\documentclass[journal]{IEEEtran}

\usepackage{amsmath,amssymb}
\usepackage[hidelinks]{hyperref}
\hypersetup{colorlinks=true,citecolor=blue,linkcolor=blue,
    filecolor=blue,urlcolor=blue}
\usepackage{cite}
\usepackage{algorithmic}
\usepackage{algorithm}

\usepackage{array}
\usepackage{graphicx}
\usepackage{caption}
\usepackage[hypcap=true,labelformat=parens]{subcaption}

\usepackage{tikz}
\tikzset{>=latex}
\usepackage{pgfplots}
\pgfplotsset{compat=newest} 
\pgfplotsset{plot coordinates/math parser=false}
\usetikzlibrary{calc, math, positioning, shapes, shapes.geometric, arrows, decorations.markings, shadows, shadows.blur} 

\usepackage{color}
\usepackage[normalem]{ulem}
\usepackage{framed}

\usepackage{makecell}
\usepackage{array}
\usepackage{multirow}
\usepackage{rotating}

\let\oldforall\forall
\renewcommand{\forall}{~\oldforall}

\usepackage{tabularx}
\newcolumntype{C}[1]{>{\centering\arraybackslash}p{#1}} 
\newcolumntype{R}[1]{>{\raggedleft\arraybackslash}p{#1}} 
\newcolumntype{L}[1]{>{\raggedright\arraybackslash}p{#1}} 

\newcommand{\fig}[1]{\figurename~\ref{#1}} 
\newcommand{\tab}[1]{Table~\ref{#1}} 
\newcommand{\sect}[1]{Section~\ref{#1}} 
\newcommand{\ilp}[1]{ILP Formulation~\ref{#1}} 

\usepackage{booktabs}
\usepackage{hyperref}
\usepackage{multirow}
\usepackage{float}

\newcommand{\AS}{\mathcal{A}} 

\newcommand{\TR}{\mathcal{T}} 

\newcommand{\aop}{\mathcal{A}}
\newcommand{\odd}{\textnormal{odd}}

\newfloat{ilp_formulation}{t}{lob}
\floatname{ilp_formulation}{ILP Formulation}

\usepackage{pdfcomment}

\begin{document}

\title{Design of Optimal Multiplierless FIR Filters}

\author{Martin~Kumm,
Anastasia~Volkova,
and Silviu-Ioan~Filip
\thanks{M.~Kumm is with the Fulda University of Applied Sciences, 36037 Fulda, Germany (e-mail: martin.kumm@cs.hs-fulda.de).}
\thanks{A.~Volkova is with University of Nantes, Nantes, France (e-mail: anastasia.volkova@univ-nantes.fr)}
\thanks{S.I.~Filip is with University of Rennes, Inria, CNRS, IRISA, Rennes, France (e-mail: silviu.filip@inria.fr).}
}


\maketitle

\begin{abstract}
This work presents two novel optimization methods based on integer linear
programming (ILP) that minimize the number of adders used to implement a 
direct/transposed finite impulse response (FIR) filter adhering to a given 
frequency specification. The proposed algorithms work by either fixing the 
number of adders used to implement the products (multiplier block adders) or
by bounding the adder depth (AD) used for these products. The latter can be 
used to design filters with minimal AD for low power applications. In contrast 
to previous multiplierless FIR approaches, the methods introduced here ensure 
adder count optimality. To demonstrate their effectiveness, we perform 
several experiments using established design problems from the literature, 
showing superior results.
\end{abstract}


%
\IEEEpeerreviewmaketitle

\section{Introduction}

\IEEEPARstart{F}IR filters are fundamental 
building blocks in digital signal processing (DSP). They provide strict 
stability and phase linearity, enabling many applications. However, their 
flexibility typically comes at the expense of a large number of 
multiplications, making them compute-intensive. Hence, many attempts
have been made in the last four decades to avoid costly multiplications and 
to implement FIR filters in a multiplierless way~\cite{lp83,s89,h96,rbd00,
gj01,ys01,gw02,v05,yl07,a08,yl09,ysl09,sy11a,sy11b,szlm12,by14,ye14,
mw15,yly17,lmyy17}.

\begin{figure}[t]
    \centering
		\subcaptionbox{SCM with $23$\label{fig:scm_example}}[3.5cm]{\includegraphics[scale=1.0]{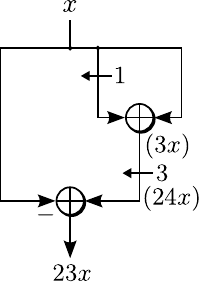}}
		\hspace{0.7cm}
		\subcaptionbox{MCM with $\{7,23\}$\label{fig:mcm_example}}[3.5cm]{\includegraphics[scale=1.0]{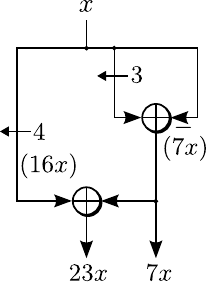}}
  \caption{Different adder circuits for constant multiplications.} 
  \label{fig:scm_mcm_example}
\end{figure}

One of the most promising ways to do so is to replace constant multiplications 
by additions, subtractions and bit shifts. Take for example the multiplication
by a constant coefficient 23. It can be computed without dedicated multipliers
as 
\begin{align}\label{eq:scm}
	23 x = 8\cdot (2x + x) - x = ((x<<1) + x)<<3) - x,
\end{align}
where $(x << b)$ denotes the arithmetic left shift of $x$ by $b$ bits. This 
computation uses one addition and one subtraction. As the add and subtract 
operations both have similar hardware cost, the total number of add/subtract 
units is usually referred to as~\textit{adder cost}. Bit shifts can be hard-wired 
in hardware implementations and do not contribute any cost. For~\eqref{eq:scm}, 
this is illustrated in~\fig{fig:scm_example}. In general, the task of finding a 
minimal adder circuit for a given constant is known as the single constant 
multiplication (SCM) problem and is already an NP-complete optimization 
problem~\cite{cs84}. 

Such a problem extends to multiplication with multiple constants,
which is necessary when implementing FIR filters. 
It is called multiple constant multiplication (MCM). Here, some of
the intermediate factors like the adder computing $3x$ in~\fig{fig:scm_example}
can be shared among different outputs. Take for example the coefficients $\{7,23\}$; 
\fig{fig:mcm_example} shows a solution for multiplying with both coefficients
at an adder cost of only two. The corresponding optimization problem is called
the MCM problem and has been addressed by numerous heuristic~\cite{dm95a,vp07,g07a}
and optimal~\cite{g08,agf10,k18} approaches.

\begin{figure}[t]
    \centering
		\subcaptionbox{Direct form\label{fig:fir_generic_direct_form}}{\includegraphics[scale=1.0]{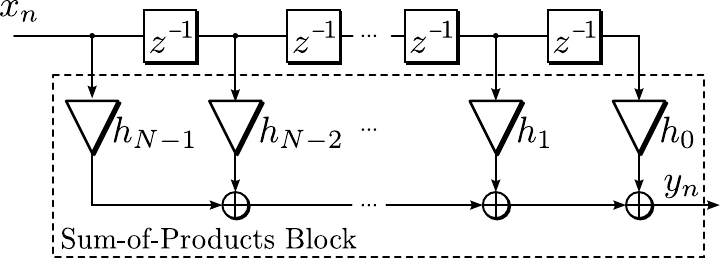}}

		\subcaptionbox{Transposed form\label{fig:fir_generic_transposed_form}}{\includegraphics[scale=1.0]{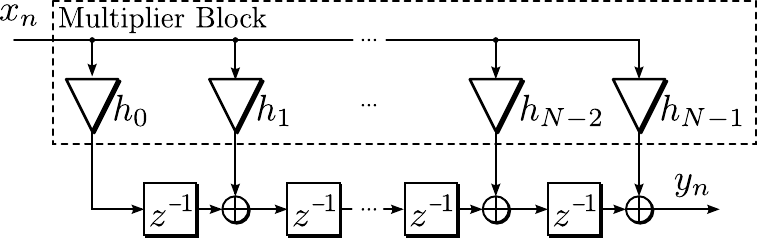}}
  \caption{Structure of FIR filters.}
  \label{fig:fir_structures}
\end{figure}

\fig{fig:fir_structures} shows the two most popular structures used
to implement FIR filters: the direct and transposed forms. The result
of an MCM solution can be directly placed in the multiplier block of
the transposed form from~\fig{fig:fir_generic_transposed_form}. The 
total adder cost is the sum of the number of~\emph{multiplier block
adders} and the remaining ones, commonly called~\textit{structural
adders}. 
The transposed form can be obtained from the direct form 
by transposition~\cite{co75}. As the transposition of a single-input-single-output 
system does not change the adder count, it leads to the same adder cost. 
So, in the end, it does not matter which one of the two considered filter
structures is actually optimized.


In the MCM optimization problem, it is assumed that the coefficients are known
and already quantized to a fixed-point (or integer) representation. 
The design of FIR filters with fixed-point coefficients and a minimum
frequency response approximation error is itself a well-known
optimization problem, going back to at least~\cite{k80}
(with subsequent extensions and improvements~\cite{de83,l90,k99,k05,k12}). 
However, it is often the case in practice that a bounded frequency response
is acceptable. In fact, there may be a large number (often hundreds or more) 
of different fixed-point coefficient sets that meet such a specification. 
Therefore, a lot of effort in fixed-point FIR filter design has gone into 
optimizing for resource use. In this context, the problem of finding a
minimal adder circuit for a given filter specification was addressed by
several authors~\cite{lp83,s89,rbd00,gj01,gw02,yl07,szlm12,yly17}.



However, to the best of our knowledge, no previous work has addressed the design
of multiplierless FIR filters in an optimal way. Here, by~\emph{optimal} multiplierless filter
we mean a direct/transposed form FIR filter requiring a minimum number of adders
to meet a target frequency specification, as well as constraints on the maximum
coefficient word size and filter order. The main contributions of our work are 
as follows:
\begin{itemize}
    \item We present for the first time a solution for the optimal multiplierless 
    design of FIR filters from a frequency specification using an ILP formulation.
	\item We provide another ILP formulation that is capable of additionally 
	limiting the adder depth inside the FIR filter.
    \item We show that relevant problem sizes can be addressed by current 
    ILP solvers and that the adder complexity of well-known FIR filters 
    can be further reduced compared to the most advanced methods.
\end{itemize}


In the following, we will give background information about previous work 
this paper is based on. In~\sect{sect:ilp_1} and~\sect{sect:ilp_2} we describe
the two ILP formulations that are at the core of the paper, whereas 
in~\sect{sec:reducing_problem_complexity} we talk about ideas meant to improve
the practical runtime of the proposed algorithms. We then present experimental 
results accompanied by a comparison with the state-of-the-art (\sect{sect:results}),
followed by concluding remarks (\sect{sect:conclusion}).


\section{Background}\label{sect:filter_design_problem}
\begin{figure}[]
    \centering
    \scalebox{0.73}{
        \begin{tikzpicture}
            \node[draw,align=center,fill=black!5, thick,rounded corners, blur shadow, minimum width=17ex, minimum height=7ex] (generator) at (-4ex,2ex) {filter coefficient \\ optimizer};
            \node[left] (bands)         at (-19ex, 12ex) {frequency bands $\Omega$};
            \node[left] (bounds)        at (-19ex, 7ex) {bounds $\underline{D}(\omega),\overline{D}(\omega)$};
            \node[left] (wordlength)    at (-19ex, 2ex) {effective word length $B$};
            \node[left] (order)         at (-19ex, -3ex) {filter order $N$};
            \node[left] (type)          at (-19ex, -8ex) {FIR filter type};

            \draw[dotted,black!50, thick] (-42ex,-14ex) rectangle +(25ex,29ex);
            \node[black!50, rounded corners]  at (-30ex,-13ex) {\emph{Input parameters}};

            \node[right] (coeffs)        at (10ex, 7ex) {scaled coefficients $h_m'$};
            \node[right] (gain)          at (10ex, 2ex) {filter gain $G$};
            \node[right] (circ)          at (10ex, -3ex) {multiplier-less solution};

            \draw[dotted,black!50, thick] (9ex,-8ex) rectangle +(24ex,20ex);
            \node[black!50, rounded corners]  at (21ex,-7ex) {\emph{Optimizer outputs}};

            \draw[->,>=stealth, thick] (bands.east) -- (generator.160);
            \draw[->,>=stealth, thick] (bounds.east) -- (generator.170);
            \draw[->,>=stealth, thick] (wordlength.east) -- (generator.180);
            \draw[->,>=stealth, thick] (order.east) -- (generator.190);
            \draw[->,>=stealth, thick] (type.east) -- (generator.200);

            \draw[->,>=stealth, thick] (generator.10) -- (coeffs.west); 
            \draw[->,>=stealth, thick] (generator.0) -- (gain.west);
            \draw[->,>=stealth, thick] (generator.350) -- (circ.west);
        \end{tikzpicture}
    }
    \caption{Simplified multiplierless FIR filter design flow.\label{fig:tool_flow}}
\end{figure}
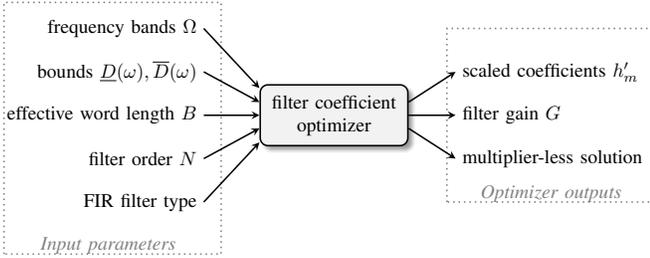

Multiplierless filter design problems usually start with a functional
specification of the frequency domain behavior, together with the number
of filter coefficients and their word lengths. An optimization procedure
is applied to get a set of bounded integer coefficients together with
their associated adder circuits for the constant multiplications needed
in the final implementation. Summarized in~\fig{fig:tool_flow}, this
section overviews these parameters and their interactions, together with
the state-of-the-art design methods found in the literature. 

\subsection{Linear Phase FIR Filters}

An $N$-th order linear phase FIR filter can be described by its zero-phase 
frequency response~\cite{Antoniou2005}
\begin{align}
    H_R(\omega) = \sum_{m=0}^{M-1} h_m c_m(\omega), && \omega \in [0,\pi],
    \label{eq:zero_phase_freq_resp}
\end{align}
which has the property that its magnitude is identical to that of the 
transfer function,~\emph{i.e.,}
\begin{align}
    |H(e^{j\omega})| = |H_R(\omega)| \ .
\end{align}
The $c_m(\omega)$ terms are trigonometric functions and $M$ denotes the 
number of independent coefficients after removing identical or negated 
ones due to symmetry. Both depend on the filter symmetry and on the parity 
of $N$ as given in~\tab{tab:filter_type_differences}~\cite{Antoniou2005}.

Let $\underline{D}(\omega)$ and $\overline{D}(\omega)$ be the desired lower 
and upper bounds of the output frequency response $H_R(\omega)$. The 
associated frequency specification-based FIR filter design 
problem consists of finding coefficients $h_m$, $m=0,\ldots, M-1$ 
that fulfill the constraints
\begin{align}\label{eq:freq_constr}
    \underline{D}(\omega) &\leqslant H_R(\omega) \leqslant \overline{D}(\omega), 
    &&\forall \omega \in \Omega,
\end{align}
where $\Omega\subseteq [0,\pi]$ is a set of target frequency bands (usually 
pass and stopbands). A standard approach in practice is to work with 
$\Omega_d\subseteq\Omega$, a uniform discretization of $\Omega$. One number 
for the size of $\Omega_d$ found in the literature is $16M$~\cite{k99}.

\subsection{Fixed-point Constraints}

Fixed-point (integer coefficient) FIR filter design problems further
restrict the search space to integer variables $h'_m \in \mathbb{Z}$ with
$|h'_m| < 2^{B}$, where the coefficients of $H_R(\omega)$ are
\begin{align}
    h_m = 2^{-B} h'_m \,
    \label{eq:h_m}
\end{align}
and $B\in\mathbb{N}$ is the~\emph{maximum effective word length} of each
coefficient (excluding sign bit).

\begin{table}[t]
  \centering
  \caption{Relation between filter order $N$, number of coefficients $M$ 
  and function $c_m(\omega)$ for different filter types}
    \label{tab:filter_type_differences}
    \begin{tabular}{lllll}
      \toprule
      Type & Sym. & $N$  & $M$ & $c_m(\omega)$\\
      \midrule
        I  & sym.  & even & $\frac{N}{2}+1$ & 
$    
c_m(\omega) =
\begin{cases}
    1                   & \text{for } m=0\\
    2 \cos(\omega m) &  \text{for } m>0 \  \\
    \end{cases}
$\\
        \cmidrule(rl){1-5}
       II  & sym.  & odd  & $\frac{N+1}{2}$ & $c_m(\omega) = 
       2\cos(\omega(m+1/2))$\\
        \cmidrule(rl){1-5}
       III  & asym. & even & $\frac{N}{2}$   & $c_m(\omega) = 
       2\sin(\omega (m-1))$\\
        \cmidrule(rl){1-5}
       IV  & asym. & odd  & $\frac{N+1}{2}$ & $c_m(\omega) = 
       2\sin(\omega(m+1/2))$\\
      \bottomrule
    \end{tabular}
\end{table}



To broaden the feasible set of efficient designs, some applications
allow the use of a real-valued scaling factor $G>0$ when computing the 
quantized fractional coefficients $h_m$. Equation~\eqref{eq:freq_constr} 
thus becomes
\begin{align}\label{eq:freq_constr_sc}
    G\underline{D}(\omega) &\leqslant H_R(\omega) \leqslant G\overline{D}(\omega),
    &&\forall \omega \in \Omega \ .
\end{align}
When the frequency specification contains a passband, it is called
the passband gain~\cite{l90}. Finding adequate bounds for $G$ is
dependent on the set/format of feasible $h_m'$ coefficient values.
If these values are constrained to a power of two space, the ratio
between the upper and the lower bound on $G$ does not need to be
larger than $2$\cite[Lemma 1]{l90}. Even when this is not the case,
the interval $[0.7, 1.4]$ is frequently used~\cite{l90,sy11a,yly17}.
For our tests, unless otherwise stated, we prefer the slightly
different interval $[2/3,4/3]$ since it is centered around one. 
In case a unity or fixed-gain filter is required we set the gain to $G=1$.

\subsection{Multiplierless FIR Filters}


Formulas~\eqref{eq:h_m} and~\eqref{eq:freq_constr_sc} are easily
expressed as constraints in an ILP formulation. However, to ensure
an optimal multiplierless design, further constraints are needed.


The way these constraints are constructed and used has varied over the years.
Early research in this direction looked at multiplierless designs where each
coefficient was represented by a limited number of signed power-of-two terms, 
optimized using branch-and-bound techniques~\cite{lp83}. 
Later, minimum signed digit (MSD) representations characterized by a
minimum number of non-zero power-of-two terms were quickly adopted for
this purpose~\cite{s89,gj01,ys01,a08,szlm12}.


MSD representations can be used to find sharing opportunities of intermediate
computations like the $7x$ term shown in~\fig{fig:mcm_example}.
One way is by searching and eliminating redundant
bit patterns common to several coefficients, a technique called common
subexpression elimination (CSE). Savings are obtained by performing the
computation specified by the bit pattern and distributing the
result to all coefficients depending on it~\cite{dm95a,h96,gw02,v05}. 
However, the CSE search cannot deliver all possible sharing opportunities
due to its dependency on the number representation~\cite{vp07} and
the effect of hidden non-zeros~\cite{fc10}. To avoid them, graph-based
approaches are commonly used in state-of-the-art MCM
methods~\cite{dm95a,vp07,g07a,g08,agf10,k18}.
Some early work on multiplierless FIR filter design already considered
this by incorporating the graph-based MCM algorithm of~\cite{dm95a} into
a genetic algorithm that optimizes the filter coefficients according to
the adder cost~\cite{rbd00}. A different approach is followed by~\cite{yl07},
where a branch-and-bound-based ILP optimization is used; here, a 
pre-specified set of integer terms, called the subexpression space, has to be 
provided that can be shared among the different coefficient expansions.  
This work was later extended with a dynamic subexpression space expansion 
algorithm~\cite{yl09,sy11a}, which, at least in the case of~\cite{sy11a}, 
claims to usually produce designs with a minimal number of adders. In contrast 
to these potentially slow branch-and-bound approaches, in~\cite{by14}, a fast
polynomial-time heuristic for the design of low complexity multiplierless
linear-phase FIR filters was proposed.

Recent work has also focused on integrating filter coefficient sparsity, 
which can also have a big impact on the complexity of the final design~\cite{ye14} 
by reducing the number of structural adders. Also, other
structures than the direct and transposed forms (see~\fig{fig:fir_structures})
have been shown to possess good properties. The factoring of FIR filters
into a cascade of relatively small subsections can lead to a lower bit-level
complexity~\cite{sy11b,mw15}. Alternative structures have also been 
proposed~\cite{lmyy17}; they provide lower word sizes for the structural 
adders.

Besides optimizing the adder count, it was shown early that the power consumption of the 
resulting filter also strongly depends on the adder depth (AD), which 
is defined as the number of cascaded adders in the multiplier 
block~\cite{ddk00,ddk02b}. 
Since then, many works have focused on limiting the AD either in MCM 
algorithms~\cite{ddk02a,fc10,k15} or directly in multiplierless filter
design methods~\cite{sy11a}. Again, all of those approaches are heuristics 
that provide minimal AD but not guarantee minimal adder cost.
Looking at the average adder depth in the structural adders can 
also help reduce power consumption~\cite{yly17}.

\section{Multiplierless FIR Filters with Fixed Number of Multiplier Block Adders}
\label{sect:ilp_1}

Our first ILP model targets the design of generic multiplierless FIR filters
regardless of their adder depth. It is based on a recently proposed MCM ILP
formulation~\cite{k18}, where the goal is to directly compute the parameters
of an MCM adder graph, if feasible, for a given number of adders. This idea is
extended here for multiplierless FIR filter design by adding constraints on the
frequency specification. As a result, we get an ILP model to design a
multiplierless filter for a fixed number of adders in the multiplier block.
To optimize the total number of adders, this ILP model is solved several times
using an overall algorithm discussed in~\sect{sect:fir_ilp1_overall}. 
In the following, we first
present the ILP formulation for the fixed number of multiplier block adders.

\subsection{ILP Formulation for Fixed Multiplier Block Adder Count}

The proposed ILP formulation is given in~\ilp{ilp:multiplierless_fir} and 
uses the constants and variables listed in~\tab{tab:constants_and_variables_of_ilp1}.
The objective is, given a fixed number of multiplier block adders $A_\text{M}$,
to minimize the number of structural adders $A_\text{S}$ (which depend
on the number of zero filter coefficients, encoded by the binary decision 
variables $h_{m,0}$).

The resulting constraints can be roughly divided into frequency response conditions
(C1, C2), equations linking the filter coefficients with the coefficients of the
multiplier block (C3) and formulas describing the multiplierless realization of the
multiplier block (C4 -- C8). 

The integer coefficients $h'_{m}$ ($m=0,\ldots,M-1$) of the FIR filter 
are directly used as integer variables in the ILP formulation. The
resulting frequency response is constrained in C1a by
setting~\eqref{eq:zero_phase_freq_resp} and~\eqref{eq:h_m}
into~\eqref{eq:freq_constr_sc}. 
Constraints C1b are so-called~\emph{lifting constraints}. These are actually 
not required to solve the problem, but can significantly reduce the search
space and improve runtime performance. Specifically, they limit the range
of the coefficients to lower $\underline{h_m}$ and upper $\overline{h_m}$
bounds. The computation of these bounds is considered
in~\sect{sec:reducing_coefficient_range}. Constraint C2 limits the range of
the gain as discussed in~\sect{sect:filter_design_problem}.

Constraints C3a to C3c provide the connection between the filter coefficient 
$h'_{m}$ and the (potentially shifted and sign-corrected) multiples computed
in the multiplier block $c_a$ or a zero coefficient. For that, the binary decision
variables $o_{a,m,s,\phi}\in\{0,1\}$ encode if $h'_{m}$ is connected to adder
$a$ of the multiplier block, shifted by $s$, and either added ($\phi=0$) or
subtracted ($\phi=1$) in the structural adders (C3a). In case the coefficient
is zero, a single binary decision variable $h_{m,0}$ is used (C3b). This encoding
allows the optimization of structural adders by considering the $h_{m,0}$ variables
in the objective function. For every zero coefficient, the corresponding structural
adder(s) can be saved depending on the coefficient and filter type.
\tab{tab:filter_type_structural_adders} shows the number of structural adders for
the different filter types. Overall, constraints C3c ensure that only one of the
above cases is valid.

\newlength{\eqspace}

\begin{ilp_formulation}
    \caption{Multiplierless FIR filters with fixed $A_\text{M}$}
	\label{ilp:multiplierless_fir}
	\setlength{\eqspace}{-2mm}
    \begin{align*} 
        \text{minimize} \ A_\text{S}(h_{m,0})
    \end{align*}
    subject to
    \begin{align*}
    &\text{C1a:}   & G 2^{B} \underline{D}(\omega) \leqslant \sum_{m=0}^{M-1} h'_{m} c_m(\omega) \leqslant G 2^{B} \overline{D}(\omega), \forall \omega \in \Omega_d     \\
    &\text{C1b:}   & \underline{h_m} \leqslant h'_{m} \leqslant \overline{h_m} , \forall m = 0, \ldots, M-1     \\
    &\text{C2:}    &  2/3 \leqslant G \leqslant 4/3\\
    &\text{C3a:}   &  h'_{m} = (-1)^\phi 2^s c_{a} \text{ if } o_{a,m,s,\phi}=1\\
    &              &      \forall a=0,\ldots, A_\text{M}, m=0,\ldots, M-1\\
    &\text{C3b:}   &  h'_{m} = 0 \text{ if } h_{m,0}=1, \forall m=0,\ldots, M-1\\
    &\text{C3c:}   & \hspace{-3mm} \sum_{a=0}^{A_\text{M}} \sum_{s=S_\text{min}}^{S_\text{max}} \sum_{\phi=0}^{1} o_{a,m,s,\phi} + h_{m,0} = 1,       \forall m=0,\ldots, M-1 \\
    &\text{C4:}    & \hspace{1.5cm} c_{0}=1 \\
    &\text{C5:}    & c_{a} = c^\text{sh,sg}_{a,\ell} + c^\text{sh,sg}_{a,\text{r}}, \forall a=1,\ldots, A_\text{M} \\
    &\text{C6a:}   & c_{a,i}                    = c_{k} \text{ if } c_{a,i,k}=1, \forall a=1,\ldots, A_\text{M}, i\in\{\ell,\text{r}\}\\
    &              &                                                            k=0,\ldots, a-1 \\
    &\text{C6b:}   & \sum_{k=1}^{a-1} c_{a,i,k} = 1,                             \forall a=1,\ldots, A_\text{M}, i\in\{\ell,\text{r}\}\\
    &\text{C7a:}   & c^\text{sh}_{a,i} = 2^s c_{a,i} \text{ if } \varphi_{a,i,s}=1\\
    &              &\forall a=1,\ldots, A_\text{M}, i\in\{\ell,\text{r}\}, s=S_\text{min},\ldots, S_\text{max}\\
    &\text{C7b:}   & \sum_{s=S_\text{min}}^{S_\text{max}-1} \varphi_{a,i,s} = 1,    \forall a=1,\ldots, A_\text{M}, i\in\{\ell,\text{r}\} \\
    &\text{C7c:}   & \varphi_{a,\ell,s}=0                      \forall s > 0\\
    &\text{C7d:}   & \varphi_{a,\ell,s}=\varphi_{a,\text{r},s} \forall s < 0 \\
    &\text{C8a:}   & \hspace{\eqspace} c^\text{sh,sg}_{a,i} = -c^\text{sh}_{a,i}  \text{ if } \phi_{a,i}=1, \forall a=1,\ldots, A_\text{M}, i\in\{\ell,\text{r}\}\\
    &\text{C8b:}   & \hspace{\eqspace} c^\text{sh,sg}_{a,i} = c^\text{sh}_{a,i} \text{ if } \phi_{a,i}=0 ,  \forall a=1,\ldots, A_\text{M}, i\in\{\ell,\text{r}\}\\
    &\text{C8c:}   & \phi_{a,\ell} + \phi_{a,\text{r}} \leqslant 1,                          \forall a=1,\ldots, A_\text{M}
    \end{align*}
\end{ilp_formulation}

\begin{table}
\caption{Used constants (top) and variables (bottom) in ILP Formulation~\ref{ilp:multiplierless_fir}}
\label{tab:constants_and_variables_of_ilp1}
\centering
\begin{tabular}{L{2.2cm}L{0.3\textwidth}}
\toprule
Constant/Variable & Meaning\\
\midrule
$A_\text{M} \in\mathbb{N}$ & Number of adders in the multiplier block\\
$M\in\mathbb{N}$ & Number of filter coefficients\\
$S_\text{min}$, $S_\text{max} \in\mathbb{Z}$ & Minimum and maximum shift\\
\cmidrule(rl){1-2}
$A_\text{S} \in\mathbb{N}$          & Number of structural adders\\
$h'_{m} \in \mathbb{Z}$             & Integer representation of filter coefficient\\
$h_{m,0} \in \{0,1\}$               & true, if coefficient $h'_{m}$ is zero\\
$c_{a}\in\mathbb{N}$                & Constant computed in adder $a$\\
$c_{a,i}\in\mathbb{N}$              & Constant of input $i\in\{\ell,\text{r}\}$ of adder $a$\\
$c^\text{sh}_{a,i}\in\mathbb{N}$    & Shifted constant of input $i\in\{\ell,\text{r}\}$ of adder $a$\\
$c^\text{sh,sg}_{a,i}\in\mathbb{N}$ & Shifted, sign corrected constant of input $i\in\{\ell,\text{r}\}$ of adder $a$\\
$\phi_{a,i}\in\{0,1\}$              & Sign of input $i\in\{\ell,\text{r}\}$ of adder $a$ (0:'$+$', $1$:'$-$')\\
$c_{a,i,k}\in\{0,1\}$               & true, if input $i$ of adder $a$ is connected to adder $k$\\
$\varphi_{a,i,s}\in\{0,1\}$         & true, if input $i$ of adder $a$ is shifted by $s$ bits\\
$o_{a,m,s,\phi}\in\{0,1\}$          & true, if coefficient $h'_{m}$ is connected to adder $a$, shifted by $s$ and sign $\phi$\\
$\underline{h_m}, \overline{h_m} \in \mathbb{Z}$ & Lower and upper bound for filter coefficient $m=0 \ldots M-1$\\
$G \in [2/3, 4/3]$  & Gain of a variable gain filter ($G=1$ when the gain is fixed)\\
\bottomrule
\end{tabular}
\end{table}

The remaining constraints C4 -- C8 are identical to the ones used for 
solving the MCM problem from~\cite{k18}. We give a brief description 
here, but refer the reader to~\cite{k18} for a more detailed presentation. 
The multiplier block input is viewed as a 
multiplication by factor one ($c_0=1$) and is defined with constraint C4.
Constraints C5 represent the actual add operation of adder $a$ and its corresponding
factor $c_a$. It is obtained by adding the shifted and possibly sign
corrected factors of its left input $c^\text{sh,sg}_{a,\ell}$ and its
right input $c^\text{sh,sg}_{a,\text{r}}$. The source of the adder 
inputs is encoded by C6a/b. Indicator constraints C6a are used to set 
the value $c_{a,i}$ of the adder input $i\in\{\ell,\text{r}\}$ to the 
actual factor when the corresponding decision variable $c_{a,i,k}$ is 
set. 
Indicator constraints are special constraints in which a binary variable controls whether or not a specified linear constraint is active. 
They are in-fact non-linear but supported by many modern ILP solvers and are also simple to linearize for other solvers (see \cite{k18}).
Constraints C6b make sure that only one source is selected.
The actual shift is constrained by C7a/b in a similar way: indicator 
constraints C7a are used to set the shifted factor $c^\text{sh}_{a,i}$ 
according to the corresponding decision variable $\varphi_{a,i,s}$. 

Constraints C7c and C7d are both optional lifting constraints used 
to reduce the search space. As the filter coefficients can be shifted
in constraint C3a, we can limit the constants of the multiplier block
to odd numbers. This allows us to use the well-known fact that odd
coefficients can be computed from odd numbers using one addition with either 
one operand left shifted while the other operand is not shifted
or both operands are right shifted by the same value~\cite[Theorem~3]{dm94}.

To support subtractions, indicator constraints C8a/b are used to set
the sign according to decision variable $\phi_{a,i}$. Finally, constraints
C8c ensure that at most one input of the adder can be nagative,
as subtracting both inputs is typically more hardware
demanding.

All of the integer variables from \tab{tab:constants_and_variables_of_ilp1} are computed from integer constants or booleans and represent integer values.
So they can be relaxed to real numbers to speed up the optimization.

\begin{table}[t]
  \centering
  \caption{Number of structural adders for the different filter types}
    \label{tab:filter_type_structural_adders}
    \begin{tabular}{ll}
      \toprule
      Type & no. of structural adders, $A_\text{S}(h_{m,0})$\\
      \midrule
        I  & $\displaystyle N-h_{0,0} - 2\sum_{m=1}^{M-1} h_{m,0}$\\
        \cmidrule(rl){1-2}
       II  & $\displaystyle N-2\sum_{m=0}^{M-1} h_{m,0}$\\
        \cmidrule(rl){1-2}
      III  & $\displaystyle N-2\sum_{m=1}^{M-1} h_{m,0}$\\
        \cmidrule(rl){1-2}
       IV  & $\displaystyle N-2\sum_{m=0}^{M-1} h_{m,0}$\\
      \bottomrule
    \end{tabular}
\end{table}

\subsection{Minimizing the Total Number of Adders}
\label{sect:fir_ilp1_overall}

As the number of adders in the multiplier block $A_\text{M}$ is fixed in \ilp{ilp:multiplierless_fir}, we need to iterate over various values $A_\text{M}$ to find the minimum number of total adders
\begin{align}
	A=A_\text{M}+A_\text{S} \ .
\end{align}
For that, we first search for a solution with minimal number of multiplier 
block adders by solving~\ilp{ilp:multiplierless_fir} for $A_\text{M}=0,1,2,\ldots$ 
until we obtain the first feasible solution. 

This solution with minimum multiplier block adders $A_\text{M,min}$ is not 
necessarily the global optimum as there might be a solution with 
$A_\text{M}>A_\text{M,min}$ and a smaller $A_\text{S}$. To account for this, 
we need a lower bound for the structural adders $A_\text{S,min}$. This 
is obtained once at the beginning of the overall algorithm by 
solving the problem for a maximally sparse FIR filter,
which we do by taking~\ilp{ilp:multiplierless_fir} where only 
the constraints C1 -- C3 are considered. 

In case the structural adders $A_\text{S}$ of solution with $A_\text{M}=A_\text{M,min}$ are not identical with $A_\text{S,min}$, 
we continue to further increment $A_\text{M}$ until we find a solution 
with $A_\text{S}=A_\text{S,min}$. This is a safe stopping point since, 
by the optimality of~\ilp{ilp:multiplierless_fir}, there is no solution 
with larger $A_\text{M}$ and smaller $A_\text{S}$. The solution with 
minimum total adders $A$ found so far is hence also globally optimal.
Typically, only a few iterations are necessary to reach this point.

\section{Multiplierless FIR Filters with Bounded Adder Depth}
\label{sect:ilp_2}

As discussed above, limiting the AD is important to reduce the power consumption of a filter.
Unfortunately, adapting~\ilp{ilp:multiplierless_fir} to limit the AD is not straightforward, 
as the topology of the adders and thus the AD is left open.
We present in this section a novel ILP model for the design of 
multiplierless FIR with limited AD which is based on a formulation that 
was initially designed for optimizing pipelined MCM (PMCM) 
circuits~\cite{kfmzm13,k15}.

In contrast to~\ilp{ilp:multiplierless_fir}, the possible coefficients
are precomputed for each adder stage $s$ and selected using binary
decision variables. The computation of the corresponding coefficient
sets is given next.

\subsection{Definition of Coefficient Sets}

We use some notation and definitions originally introduced in~\cite{vp07}.
First, we define the generalized add operation called $\aop$-operation, 
which includes the shifts. An $\aop$-operation has two input coefficients
$u, v \in \mathbb{N}$ and computes 
\begin{align}
    \aop_q(u,v)=|2^{l_u} u +(-1)^{s_v} 2^{l_v} v| 2^{-r},
    \label{eq:a_op}
\end{align}
where $q=(l_u,l_v,r,s_v)$ is a configuration vector which determines the left
shifts $l_u, \ l_v \in \mathbb{N}_0$ of the inputs, $r \in \mathbb{N}_0$
is the output right-shift and $s_v\in\{0,1\}$ is a sign bit which denotes
whether an addition or subtraction is performed. 

Next, we define the set $\aop_*(u,v)$ containing all possible coefficients
which can be obtained from $u$ and $v$ by using exactly one $\aop$-operation:
\begin{align}
    \label{eq:aop_star}
    \aop_*(u,v) := \{\aop_q(u,v) \ | \ q \text{ is a valid configuration}\} \ .
\end{align}
A~\textit{valid} configuration is a combination of $l_u$, $l_v$, $r$ and $s_v$
such that the result is a positive odd integer $\aop_q(u,v) \leqslant c_{\max}$.
The reason for limiting the integers to odd values is that we can compute every 
even multiple by shifting the corresponding odd multiple to the left.
The $c_{\max}$ limit is used to keep $\aop_*(u,v)$ finite. 
It is chosen as a power-of-two value which 
is usually set to the maximum coefficient bit width
$B$ plus one~\cite{vp07,dm95a}
\begin{align}
	c_{\max} &:= 2^{B+1} \ .
	\label{eq:cmax}
\end{align}
For convenience, the $\aop_*$ set is also defined for an input set
$X\subseteq\mathbb{N}$ as
\begin{align}
    \aop_*(X):=\bigcup_{u,v \in X} \aop_*(u,v) \ .
    \label{eq:aop_star_single_set}
\end{align}

We can now define the coefficients that can be computed at
adder stage $s$, denoted as $\AS^s$, by recursively computing the
$\aop_*$ sets
\begin{align}
    \AS^0 & := \{1\}\\
    \AS^s & := \aop_*(\AS^{s-1}) \ .
\end{align}
In addition, let $\TR^{s}$ denote the set of $(u,v,w)$~\textit{triplets}
for which $w \in \AS^{s}$ can be computed using $u$ and $v$ from the previous
stage (\emph{i.e.,} $u,v\in \AS^{s-1}$). $\TR^{s}$ can be computed
recursively, starting from the last stage $s$, which is equal to the maximum
allowable AD:
\begin{align}
    \TR^{s} := \{(u,v,w)\ |& \ w = \aop_{q}(u,v), \notag \\
                           & u,v \in \AS^{s}, \ u \leqslant v, \ w\in \AS^{s+1}\}.
\end{align}

To give an example, the first elements of $\TR^{1}$ are
$\TR^1 = \{(1,1,1),(1,1,3),(1,1,5),(1,1,7),(1,1,9),(1,1,15),\ldots\}$. This set
contains all the possible rules for computing multiples from 
the input within one stage of additions, while set
$\TR^2 = \TR^1 \cup \{(1,3,11),(1,5,11),\ldots,(3,5,11),\ldots\}$
contains all the combinations of how elements in the next stage can 
be computed.

\subsection{ILP Formulation for Fixed Adder Depth}

The bounded AD model is given in~\ilp{ilp:multiplierless_fir_limited_depth}, while the corresponding constants and
variables are given in~\tab{tab:constants_and_variables_of_ilp2}.

In contrast to~\ilp{ilp:multiplierless_fir}, the objective is to directly 
minimize the total number of adders $A$, which is separated into adders in the
multiplier block ($A_\text{M}$) and structural adders $A_\text{S}$.
Similar to~\ilp{ilp:multiplierless_fir}, the constraints are divided
into frequency response conditions (C1, C2), the link between the filter 
coefficients and the coefficients of the multiplier block
(C3, C4) and the equations describing the multiplierless realization of
the multiplier block (C5 -- C8).



\begin{ilp_formulation}
	\caption{Multiplierless FIR filters with depth limit}
	\label{ilp:multiplierless_fir_limited_depth}
	\setlength{\eqspace}{-12mm}
    \begin{align*} 
        \text{minimize} \ \underbrace{\sum_{s=1}^{S} \sum_{w \in \AS^s} a_w^s}_{=A_\text{M}} + A_\text{S}(h_{m,0})
    \end{align*}
    subject to
    \begin{align*}
        &\text{C1a:} & \hspace{\eqspace} G 2^{B} \underline{D}(\omega) \leqslant \sum_{m=0}^{M-1} h'_{m} c_m(\omega) \leqslant G  2^{B} \overline{D}(\omega), \forall \omega \in \Omega_d     \\
        &\text{C1b:} & \underline{h_m} \leqslant h'_{m} \leqslant \overline{h_m}, \forall m = 0,\ldots, M-1     \\
	    &\text{C2:}    &  2/3 \leqslant G \leqslant 4/3\\
        &\text{C3a:} & h'_{m} = 
        \begin{cases}
        \displaystyle \sum_{w=0}^{2^B-1} w h_{m,w} &\text{ if } \phi_m=0\\
        \displaystyle -\sum_{w=1}^{2^B-1} w h_{m,w} &\text{ if } \phi_m=1\\
        \end{cases} \\
        && \forall m=0,\ldots, M-1\\
        &\text{C3b:} & \sum_{w=0}^{2^B-1} h_{m,w} = 1, \forall m = 0,\ldots, M-1     \\
        &\text{C4:} & \hspace{\eqspace} r_{\odd(w)}^S + a_{\odd(w)}^S \geqslant \frac{1}{M} \sum_{m=0}^{M-1} h_{m,w}, \forall w=0,\ldots, 2^B-1\\
        &\text{C5:} & \hspace{\eqspace} r_w^s = 0 \forall w \in \AS^{s} \setminus \bigcup_{s'=0}^{s-1} \AS^{s'} \text{ with } s=1,\ldots, S-1\\
        &\text{C6:} & r_w^s - a_w^{s-1} - r_w^{s-1} \leqslant 0, \forall w \in \AS^s \setminus \{0\} \text{, } s=2,\ldots, S\\
        &\text{C7:} & a_w^s - \hspace{-5pt} \sum_{(u,v,w') \in \TR^s \, | \, w'=w} \hspace{-5pt}  x_{(u,v)}^{s-1} \leqslant 0\\
        && \forall w\in \AS^s, s=2,\ldots, S\\
        &\text{C8:} & 
    \begin{aligned}
        x_{(u,v)}^s - r_u^{s} - a_u^{s} \leqslant 0\\
            x_{(u,v)}^s - r_v^{s} - a_v^{s} \leqslant 0\\
    \end{aligned}
    \\ 
    && \forall (u,v,w) \in \TR^s \text{ with } s=1,\ldots, S-1
    \end{align*}
\end{ilp_formulation}

Constraints C1a/b and C2 are identical to the ones in \ilp{ilp:multiplierless_fir}.
Now, the connection between the odd multiplier block coefficients of the pre-computed sets and the filter coefficients is performed using binary decision variables.
Let $h_{m,w} \in \{0,1\}$ be a binary decision variable that is true if the magnitude of $h'_{m}$ is identical to $w$, \emph{i.e.,}
\begin{align}
    h_{m,w} = 
    \begin{cases}
    1 & \text{when } |h'_m| = w\\
    0 & \text{otherwise} \\
    \end{cases}
\end{align}
for $m=0,\ldots, M-1$ and $w=0,\ldots, 2^B-1$.
Furthermore, let $\phi_m$ determine the sign of $h'_{m}$ as follows
\begin{align}
    \phi_m = 
    \begin{cases}
    0 & \text{when } h'_m \geq 0\\
    1 & \text{otherwise} \ . \\
    \end{cases}
\end{align}

The value of each integer coefficient $h'_{m}$ is selected by the indicator constraints C3a.
In addition, constraints C3b make sure that only one value per filter coefficient is selected.

Next, we distinguish between coefficients that are computed for the selected stage (by using an addition) and coefficients that are just replicated from a previous stage.
This replication can be either implemented by a simple wire (as this was implied in \ilp{ilp:multiplierless_fir}) or in case of a pipelined implementation of the multiplier block, it will be implemented by a register. This allows to also model the register cost in the latter case (not treated here but it is a trivial extension of the objective).
Hence, we introduce two new decision variables for each $w$ and stage: $a_w^s$ and $r_w^s$, which are true, if $w$ in stage $s$ is realized using an adder or register/wire, respectively.

The connection to the filter coefficients $h_{m,w}$ is made through C4. As several of the $M$ $h_m$ coefficients can have the same $w$ value, the right hand side of C4 is scaled by $1/M$ to keep it less than one. 
Whenever the right hand side of C4 is non-zero it forces the realization of coefficient $w$ in the output stage $S$, either as an adder or as a register/wire.

\begin{table}
\caption{Used constants (top) and variables (bottom) in ILP Formulation~\ref{ilp:multiplierless_fir_limited_depth}}
\label{tab:constants_and_variables_of_ilp2}
\centering
\begin{tabular}{L{2cm}L{0.32\textwidth}}
\toprule
Constant/Variable & Meaning\\
\midrule
$M\in\mathbb{N}$                 & Number of filter coefficients\\
$\AS^s \subseteq \mathbb{N}$     & Coefficients that can be computed in adder stage $s$\\
$\TR^s \subseteq \mathbb{N}^3$   & Tripplets $(u,v,w)$ from which $w \in \AS^{s}$ can be computed using $u,v\in \AS^{s-1}$\\
$\underline{h_m}, \overline{h_m} \in \mathbb{Z}$ & Lower and upper bound for filter coefficient $m=0,\ldots, M-1$\\
\cmidrule(rl){1-2}
$h'_{m} \in \mathbb{Z}$          & Value of filter coefficient $m=0,\ldots, M-1$\\
$h_{m,w} \in \{0,1\}$              & true, if $|h'_{m}| = w$ for $m=0,\ldots, M-1$ and $w=0,\ldots, 2^B-1$\\
$\phi_{m} \in \{0,1\}$             & true, if $h'_{m}$ is negative\\
$a_w^s \in \{0,1\}$                & true, if $w \in \AS^s$ in stage $s=1,\ldots, S-1$ is realized using an adder\\
$r_w^s \in \{0,1\}$                & true, if $w \in \AS^s$ in stage $s=1,\ldots, S-1$ is realized using a register or wire\\
$x^s_{(u,v)} \in \{0,1\}$          & true, if $u$ and $v$ are available in stage $s=1,\ldots, S-2$\\
$G \in [2/3, 4/3]$ & Gain of a variable gain filter ($G=1$ when the gain is fixed)\\
\bottomrule
\end{tabular}
\end{table}

Constraints C5 and C6 consider the realization as register/wire: they require that a value $w$ can only be replicated from a previous stage if it was computed or replicated before.

The realization as an adder computing constant $w$ from the inputs $u$ and $v$ requires the presence of both inputs in the previous stage.
For that, the binary variables $x_{(u,v)}^s$ are introduced which determine if both are available in stage $s$:
\begin{align} 
  x_{(u,v)}^s =
	\begin{cases}
        1 & \text{if both $u$ and $v$ are available in stage $s$}\\
        0 & \text{otherwise}
	\end{cases}
  \label{eq:x_u_v}
\end{align}
Now, constraint C7 specifies that if $w$ is computed by $w = \aop(u,v)$ in stage $s$, the pair $(u,v)$ has to be available in the previous stage. 
If a pair ($u,v$) is required in stage $s$, constraints C8 make sure that $u$ and $v$ have been realized in the previous stage either as register or adder.

Note that instead of using constraint C5 it is more practical to remove all variables $r_w^s$ which are zero from the cost function and their related constraints.
Also note that the binary variables $x^s_{(u,v)}$ and the integer variables $h_m$ can be relaxed to real numbers to speed up the optimization.



\subsection{Selecting the Adder Depth}

The AD is often selected to be as small as possible, 
typically at the expense of a higher adder cost. It is well known that the 
minimal AD needed when multiplying with a given 
coefficient can be realized by using a binary tree~\cite{kp01}. Therefore, it cannot be lower than the base two logarithm 
of the non-zero digit count of its MSD representation. Unfortunately, as 
the coefficients are not known in advance, the minimum AD cannot be derived 
from the filter specification. However, the upper bound of the AD can be 
computed from the coefficient word size $B$ as follows. A $B$\,bit binary 
number can have up to $B+1$ digits when represented as an MSD number 
and up to $\lfloor{(B+1)/2}\rfloor+1$ non-zeros in the worst 
case~\cite{k15}. This leads to a maximum adder depth of
\begin{align}
	\text{AD}_{\max}=\log_2\left(\left\lfloor \frac{B+1}{2} \right\rfloor+1\right) \ .
	\label{eq:max_adder_depth_rel_to_word_size}
\end{align}
Using this bound, a search from $\text{AD}=0,\ldots, \text{AD}_{\max}$ can 
be performed until the first feasible solution is found. 

For practical FIR filters, early studies have shown that 
coefficient word sizes between 15\,bit to 20\,bit are sufficient to achieve 
approximation errors between $-70$ and $-100$\,dB~\cite{cr73}.
Using~\eqref{eq:max_adder_depth_rel_to_word_size}, this translates to ADs 
of at most three to four. In our experiments, we found 
very good solutions with $\text{AD}=2$ for most of the 
filters from practice.

\section{Reducing the Problem Complexity}
\label{sec:reducing_problem_complexity}

\subsection{Reducing the Coefficient Range}
\label{sec:reducing_coefficient_range}

Following~\cite[Sec.~4]{gj01}, we bound the search space for 
the coefficient values by projecting the polytope corresponding 
to the discretized versions of~\eqref{eq:freq_constr} 
or~\eqref{eq:freq_constr_sc} onto each $h_m'$. The goal is a 
tight interval enclosure $[\underline{h_m}, \overline{h_m}]$ for 
the feasible values of $h_m'$. This corresponds to the LPs:
\[ 
    \text{minimize} \ h_m' 
\]
or
\[
    \text{maximize} \ h_m'
\]
subject to
\[
	G \underline{D}(\omega) \leqslant \sum_{k=0}^{M-1} h_{k}' c_k(\omega) \leqslant G \overline{D}(\omega), \quad \forall \omega \in \Omega_d,
\]
where $h_{k}' \in \mathbb{R}$ for $k=0,\ldots, M-1$ and $G \in \left[2/3, 4/3\right]$ 
(or $G=1$ when unity gain is used).
We get $[\underline{h_m},\overline{h_m}]$ by taking
\begin{align*}
\underline{h_m} &= \left\lceil h_m' \right\rceil \text{ from minimize } h_m', \\
\overline{h_m} &= \left\lfloor h_m' \right\rfloor \text{ from maximize } h_m'.
\end{align*}

\subsection{Discretizing the Frequency Domain}\label{sec:afp}

Even though $\Omega$ is replaced by a finite set $\Omega_d$, we 
perform a rigorous posteriori validation of the result over
$\Omega$~\cite{Volk17c}. 
Still, the typically large size of $\Omega_d$ ($16M$ is a 
common value found in the literature) can have a 
big impact on the runtime of the filter design routine. 
This is shown for instance in the context of an optimal 
branch-and-bound algorithm for FIR filter design with fixed-point 
coefficients~\cite[Table 2]{k99}. A too small number of points 
can, on the other hand, lead to an invalid solution over 
$\Omega$ and a larger feasible set, potentially incurring a 
larger runtime as well.


 
It is thus important to consider a discretization of reasonable 
size that is unlikely to lead to invalid solutions over $\Omega$ 
(\emph{i.e.,} equations~\eqref{eq:freq_constr} or~\eqref{eq:freq_constr_sc} 
do not hold) and does not increase the search space by a too large factor.
To this effect, we use so-called approximate Fekete
points (AFPs), which contain the most critical frequencies for a 
given filter that needs to fit a target frequency response.
They have recently been used to improve the robustness of
the classic Parks-McClellan Chebyshev FIR filter design 
algorithm~\cite{f16} and for a fast and efficient heuristic for FIR 
fixed-point coefficient optimization~\cite{bfh18}. They are efficient  
choices when performing polynomial interpolation/approximation on 
domains such as $\Omega$. This is relevant in our context since
$H_R(\omega)$ in~\eqref{eq:zero_phase_freq_resp} is a polynomial in 
$\cos(\omega)$~\cite[Ch.~7.7]{os14}. For details on how to
compute them we refer the reader to~\cite{f16,bfh18} and the references
therein.

\subsection{An Adaptive Search Strategy}

Even if the current $\Omega_d$ leads to a solution that does not
pass a posteriori validation, it might still be possible to 
rescale the gain factor $G$ such that~\eqref{eq:freq_constr_sc} holds.  
By taking a point $\omega_{\max}\in\Omega$ where 
$G\underline{D}(\omega_{\max})-H_R(\omega_{\max})$ or
$H_R(\omega_{\max})-G\overline{D}(\omega_{\max})$ is largest
(\emph{i.e.,} the point of largest deviation from the specification) we
first update $G$ to take a value close to 
$H_R(\omega_{\max})/\underline{D}(\omega_{\max})$ or
$H_R(\omega_{\max})/\overline{D}(\omega_{\max})$, depending
on where the deviation occurs. If this new gain leads to a valid
solution over $\Omega$, then it is optimal. If not, we update $\Omega_d$
by adding the points of largest deviation for each frequency subdomain.
We rerun the optimization with this new $\Omega_d$, repeating
until either (a)
there are no more invalid frequency points or (b) the problem becomes
infeasible, meaning no solution with the imposed constraints over
$\Omega$ exists. 

We should mention that running the result validation code 
of~\cite{Volk17c} at each iteration of the adaptive routine is 
computationally expensive. This is why at each iteration we perform 
a fast, non rigurous test consisting of verifying~\eqref{eq:freq_constr_sc}
on a much denser discretization of $\Omega$ than $\Omega_d$. We found this 
to usually be sufficient in ensuring that the a posteriori 
validation~\cite{Volk17c} done at the end of optimization is successful.

\section{Experimental Results}\label{sect:results}

To test the ILP formulations discussed above, we have implemented them in a C++
filter design tool\footnote{Available as an open-source project at:
\href{https://gitlab.com/filteropt/firopt}{https://gitlab.com/filteropt/firopt}.}.
It features a flexible command-line interface.

\subsection{Experimental Setup and Parameter Choices}

All experiments were run on a Linux machine with an Intel Xeon E5-2690 v4 CPU
with 56 cores and 252 GB of RAM. 
The proposed implementation supports several popular open source
and commercial (M)ILP solvers, such as SCIP~\cite{scip}, Gurobi~\cite{gurobi}
and CPLEX~\cite{cplex}\footnote{Free academic licenses for Gurobi 8.1
and CPLEX 12.6 are used.}. For convenience, these solvers are
accessed through the ScaLP~\cite{sskz18} library, which acts as a
frontend. Based on our experiments, Gurobi usually proved to be
the fastest backend, which is why, apart from a few exceptions, use
it on all the examples below. 




\begin{figure}[t]
    \centering
    \includegraphics[width=\linewidth]{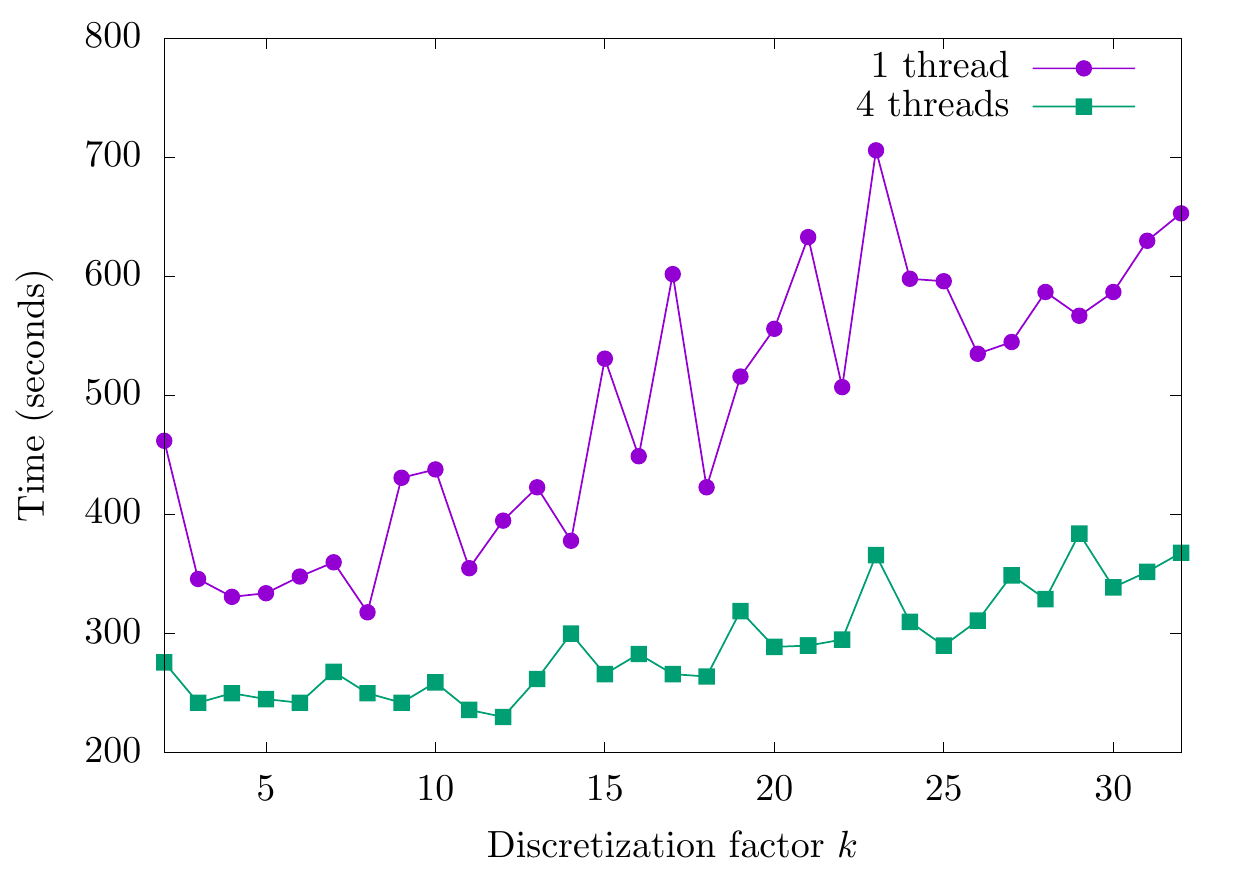}
    \caption{Total execution times for the design of a family of filters
    from Section~\ref{sec:Benchmark_Ex1} with respect to the size of
    $\Omega_d$ ($kM$ points). Each data point corresponds to the time
    for the design of 42 filters of increasing complexity.
}\label{fig:density_timing_comp}
\end{figure}

All experiments use the AFP-based frequency grid discretization mentioned
in Section~\ref{sec:afp}. 
As discussed before, the number of frequency points in $\Omega_d$ is run-time critical.
To determine an appropriate size, we ran an experiment 
using a typical design scenario with an $\Omega_d$ 
size of $kM$ points and $k=1,\ldots,32$. \fig{fig:density_timing_comp} shows the runtimes.
Not surprisingly, they start large for very low $k$, as in these cases the frequency grid 
usually has to be extended to address violations, which require re-running the optimization routine
on a new grid. As soon as $k$ is large enough (around $k\geqslant 4$), invalid results become rare, 
meaning just one optimization pass is sufficient. Further increasing $k$ at this point just leads
to more constraints in the model and likely a larger runtime for the optimizer.
Based on these results, we selected to start with $4M$ points. We find this choice 
usually delivers a good balance between optimizer runtime and number of iterations needed 
to obtain a valid solution over $\Omega$.

\begin{table}[t]
	\caption{Specifications of the reference filters}\label{tab:spec_reference_filters}
	\begin{subtable}{\linewidth}
		\caption{Specifications S1, S2, L1, L2 and their variations. }				
		\centering 
		\begin{tabular}{@{}llllll@{}}
			
			\toprule
			Name & Source & $\Omega_p/\pi$ & $\Omega_s/\pi$ & $\delta_p$ & $\delta_s$ 
			\\ \midrule
			S1a & \cite{s89}          &   $[0, 0.3]$         &          $[0.5, 1]$  &   $0.00645$         &  $0.00645$     
			\\ \hline
			S1b & \cite{rbd00}          &   $[0, 0.3]$         &          $[0.5, 1]$  &   $0.00636$         &  $0.00636$     
			\\ \hline
			S1c & \cite{yl07, sy11a}          &   $[0, 0.3]$         &          $[0.5, 1]$  &   $0.01570$         &  $0.00660$        \\ \hline
			S2a & \cite{s89, a08}          &     $[0, 0.042]$         &          $[0.14, 1]$   &   $0.01160$         &  $0.001$   
			\\  \hline
			S2b & \cite{yl07, sy11a}          &   $[0, 0.042]$         &          $[0.14, 1]$  &   $0.01200$         &  $0.001$ 
			\\  \hline
			L1 & \cite{a08, yl07} & $[0.8, 1]$ & $[0, 0.74]$ & $0.0057$ & $0.0001$ \\ \hline 
			L2 &  \cite{lp83, szlm12}         &    $[0, 0.2]$        &  $[0.28, 1]$           &  $0.02800$          &   $0.001$     \\ 
			\bottomrule
		\end{tabular}
	\end{subtable}\\
	
	\vspace{\baselineskip}
	\begin{subtable}{\linewidth}
		\centering
		\caption{Specification of the multiband filter L3.}	
		\begin{tabular}{@{}lll@{}}
			
			\toprule
			$\Omega_i / \pi$ & $\underline{D}(\omega)$ & $\overline{D}(\omega)$ \\ \midrule
			$[0, 0.15]$ &       $0.9772$                  &        $1.0232$                \\
			$[0.15, 0.1875]$&     $0.9441$                    &       $1.0232$                 \\
			$[0.1875, 0.2125]$ &    $0.9016$                     &      $1.0232$                  \\
			$[0.2875, 1]$ & $0$ & $0.0316$ \\
			\bottomrule
		\end{tabular}
	\end{subtable}
\end{table}

\subsection{Benchmark Set}

Several multiplierless filter designs were computed
to evaluate our methods. They are introduced next.

\subsubsection{A Family of Specifications from~\cite[Example 1]{rbd00}}\label{sec:Benchmark_Ex1}

We consider a family of low-pass linear-phase filter specifications
from Redmill et al.~\cite{rbd00}. 
These specifications are defined by:
	\begin{align*}
1 - \delta &\leqslant H_R(\omega) \leqslant 1 + \delta,  &\omega \in [0, 0.3] \quad \text{(passband)} \\
-\delta &\leqslant H_R(\omega) \leqslant \delta,  &\omega \in [0.5, 0.1] \quad \text{(stopband)}
\end{align*}
where $\delta$ is a parameter regulating error. We set
${\delta=10^{-\frac{p}{20}}}$, where $p>0$ is the error in decibels (dB).

Our goal with this benchmark is to explore the tradeoff between the error ($p$),
the filter order ($N$) and the word length ($B$) in terms of the total
number of adders.

\subsubsection{A Set of State-of-the-art Specifications}
We also test our tool on a set of reference specifications from the literature~\cite{lp83,s89,rbd00,yl07,ysl09,sy11a,szlm12}, referred to 
as $S1$, $S2$, $L1$, $L2$ and $L3$. They 
are all low-pass filters defined by
\begin{align*}
1 - \delta_p &\leqslant H_R(\omega) \leqslant 1 + \delta_p,  &\omega \in \Omega_p \quad \text{(passband)}, \\
-\delta_s &\leqslant H_R(\omega) \leqslant \delta_s,  &\omega \in \Omega_s \quad \text{(stopband)},
\end{align*}
where the values of $\delta_p, \delta_s, \Omega_p, \Omega_s$ for each specification
are given in Table~\ref{tab:spec_reference_filters}. Over time, these reference filter
specifications were slightly modified by different publications. To compare with each
one, we indicate variations by suffixes, \emph{e.g.} S1a and S1b.

We note that this restriction to low-pass filters comes only from the
existing literature and that our tool can be successfully used for the design of
other types of filters, such as multiband filters or decimators (since we generalize
constraints on the frequency response as~\textit{functions} of frequencies).

\subsection{Results}

\subsubsection{ILP Formulation 1 vs. ILP Formulation 2}\label{sec:ILP1vsILP2}

In the first experiment, we compare~\ilp{ilp:multiplierless_fir} (in its
overall form discussed in Section~\ref{sect:fir_ilp1_overall})
to~\ilp{ilp:multiplierless_fir_limited_depth}.
Both models can be used to optimize for the total number of adders
(MB and structural) given fixed parameters like filter order $N$, filter type 
and the effective word length $B$ (see \fig{fig:tool_flow}). 
In case of ILP2, the adder depth is an additional constraint.
Therefore, in practice, the two approaches can sometimes lead to different results.

\begin{figure}[t]
	\centering
	\includegraphics[width=\linewidth]{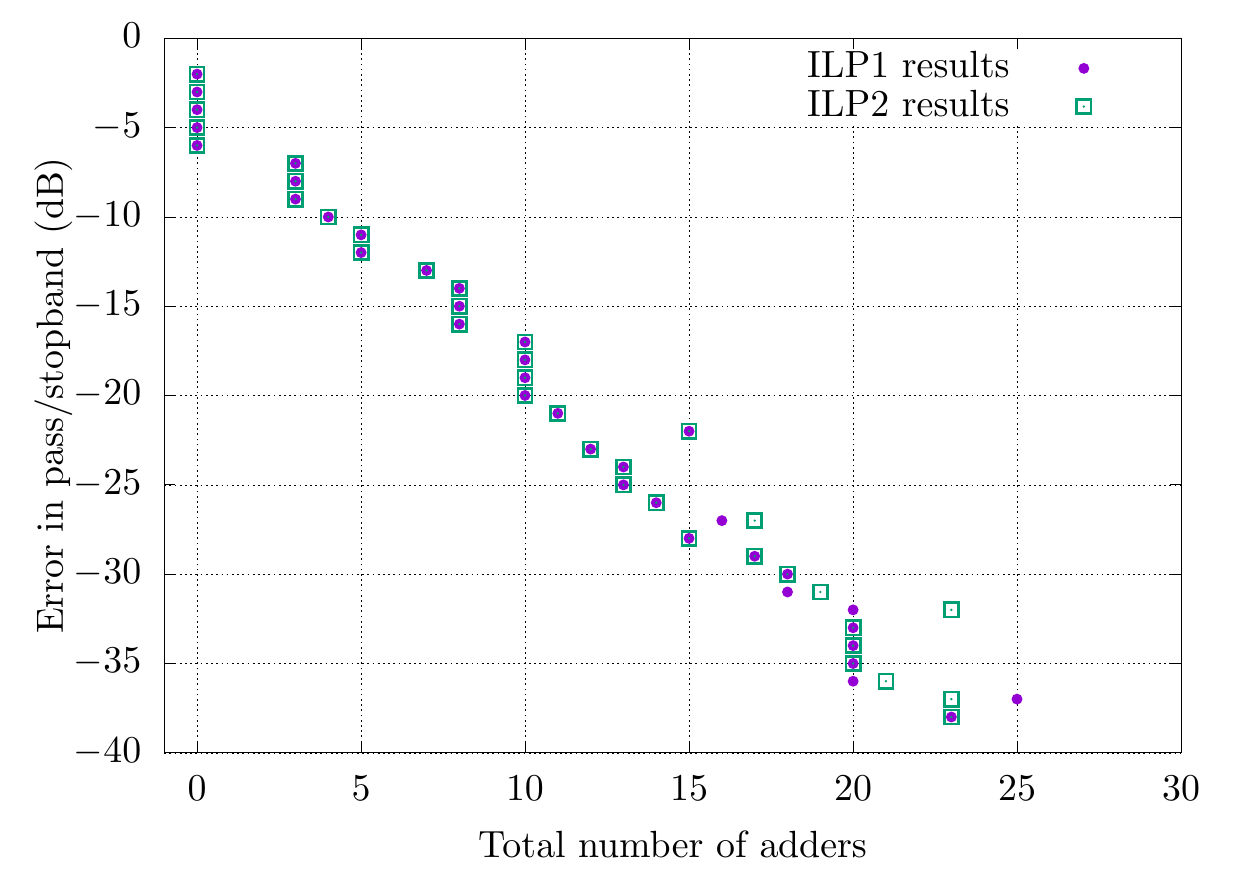}
    \caption{Total adder count comparison when using ILP1 (adapted to 
    minimize the total number of adders) and ILP2 (with AD limit set 
    to 2) on the Redmill set of filters~\cite{rbd00}.}
    \label{fig:ILP1_ILP2}
\end{figure}

This is exemplified in~\fig{fig:ILP1_ILP2}, where we design a set of 
filters using the family of specifications from~\cite[Example 1]{rbd00} as 
described in~\sect{sec:Benchmark_Ex1}.
We consider $2\times 37$ filters corresponding to $p=2,\ldots,38$ 
(\emph{i.e.,}~error is varied from $-2$\,dB to $-38$\,dB), with a 
$9$-bit effective word length and fixed gain $G=1$. In each case, a type I 
filter with smallest $N$ that leads to a feasible solution under the given 
constraints was used. For ILP2, the upper bound on the AD is set to $2$ as 
this turned out to be sufficient for all test instances.

For most error targets the resulting total adder count is identical between 
the two. The exceptions are $-27, -31, -32$ and $-36$\,dB, where ILP1 gives 
a better total adder count, and $-37$\,dB, where ILP2 is better.
The four cases where ILP1 gives a better result are not surprising considering
that the limited AD in ILP2 restricts the coefficient search space. 
For $-37$\,dB, the difference comes from the fact that the ILP1 solver 
is able to find an optimal solution with $N=20$, while an $\text{AD}=2$ 
solution for ILP2 is only possible starting with $N=22$. Taking $\text{AD}=3$ 
with ILP2 gives the $N=20$ solution found with ILP1.

Due to the different nature of the constraints and objective values of both ILP
models, it is hard to do a runtime comparison between the two. We only mention that
when there is a feasible filter with small number of MB adders, ILP1 can be quite
fast for moderate size problems ($N<50$), but turns out to struggle for
problems with many MB adders. Similarly, for feasible designs with small 
AD (up to 3), ILP2 will be fast for $N<50$ and overall scales better than ILP1.

In the rest of the paper, for comparison with previous work, we 
use~\ilp{ilp:multiplierless_fir_limited_depth} with $\text{AD}=2$ (unless otherwise stated).


\subsubsection{Design Space Exploration for~\cite[Example 1]{rbd00}}\label{sec:redmill_results}
\begin{figure}[t]
    \centering
    \includegraphics[width=\linewidth]{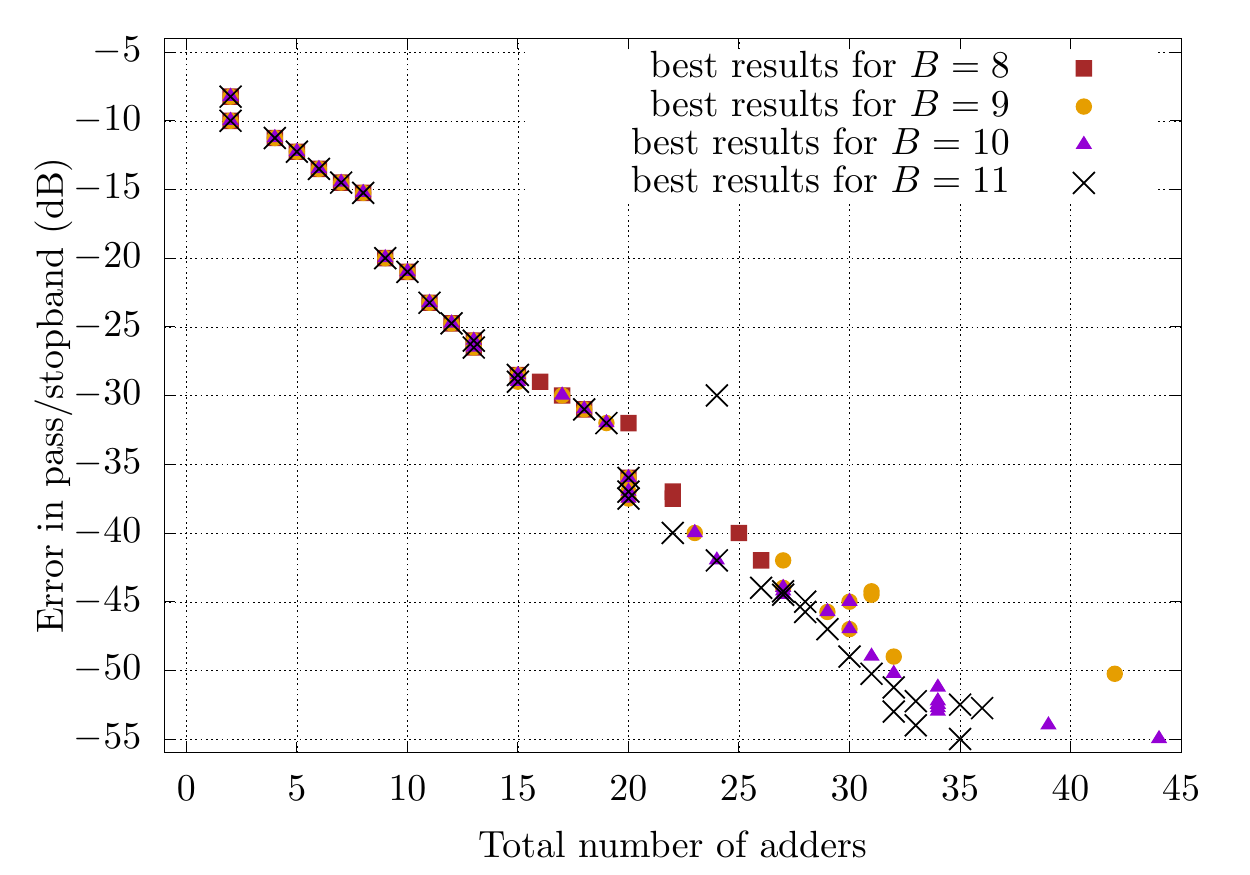}
    \caption{Our designs with effective word lengths varying from 8 to 11, filter 
    type I and smallest feasible filter order.}\label{fig:Redmill_wordlengths}
\end{figure}

The experiment setting from~\sect{sec:ILP1vsILP2} is expanded upon. We
compare our best results (with effective word lengths $B\in\{8,9,10,11\}$)
with those from~\cite[Example 1]{rbd00}. {We start off by considering
only type I filters (just like in~\cite{rbd00}), variable gain $G\in[2/3,4/3]$
and minimal order $N$ for each error target. The results are illustrated
in~\fig{fig:Redmill_wordlengths}. 
We note that there are certain cases where, for a given $B$, 
taking the minimal filter order leading to a feasible solution \emph{does not}
minimize the adder cost. This is most visible for $B=11$ and a $-30$dB error
target, where a minimal order $N=14$ filter requires $24$ adders. For $B=10$, the
minimal $N$ is $16$, leading to only $17$ total adders, a $7$ adder improvement.
Taking $N=16$ for $B=11$ also results in a $17$ adder solution.} 
A lower implementation cost is sometimes possible when increasing the filter 
order leads to a sparser filter and/or a more economical MCM design. Such 
solutions better optimize the objective functions in the proposed ILP models. 
We nevertheless remark that increasing the filter order beyond a certain 
threshold will not lead to different solutions~\cite{k13}.

Of course, increasing the
word length can also lead to a significant improvement in the results. For
instance, the optimal $-50$dB atteanuation results for $B\in\{9, 10, 11\}$ require
$41, 32$ and $31$ adders, respectively.

\begin{figure}
    \includegraphics[width=\linewidth]{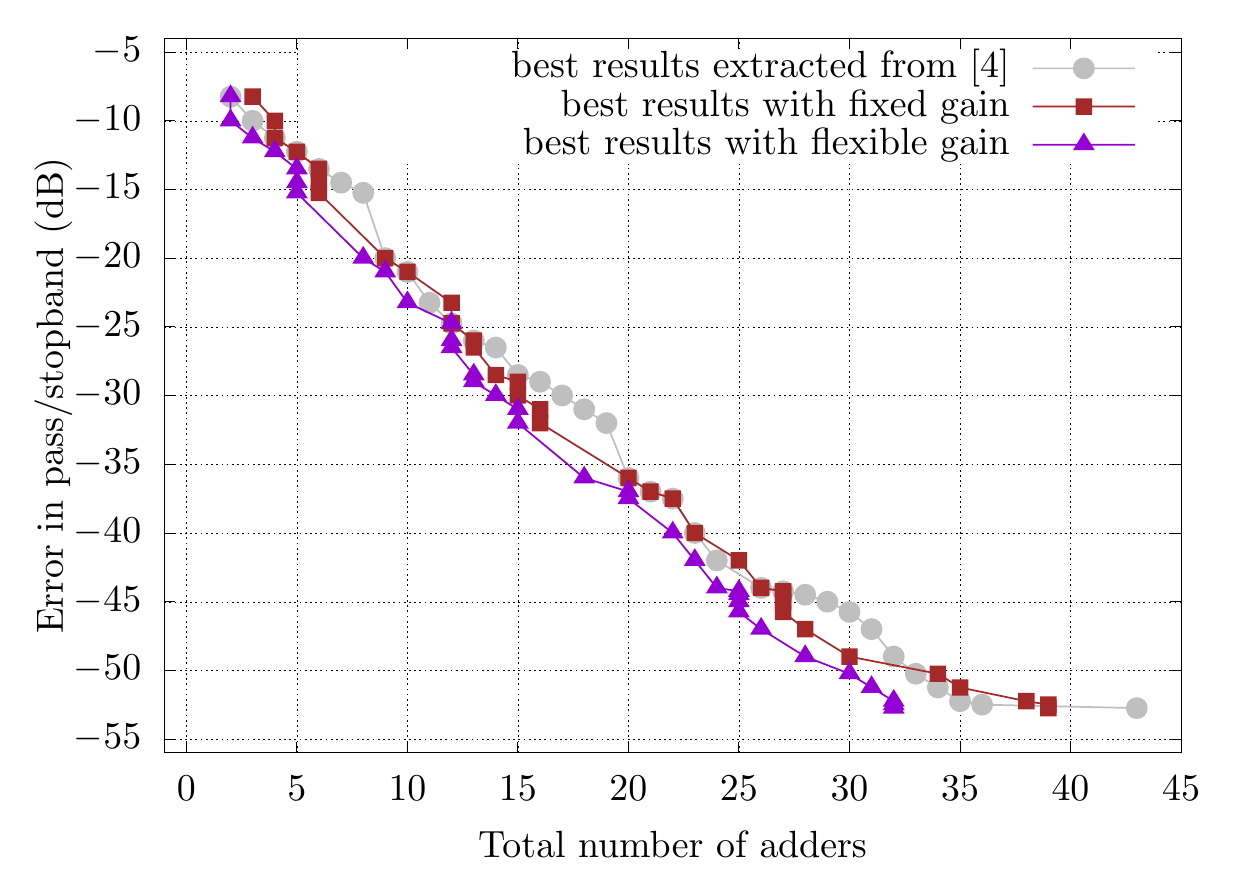}
    \caption{Comparison between our best design space exploration results and the best results from~\cite[Example 1]{rbd00}.
    Our tool improves designs from~\cite{rbd00} or proves them optimal.}\label{fig:Redmill}
\end{figure}

This nonlinearity of the word length/cost relation means that the user should favor
a comprehensive exploration of the design space, varying the design parameters 
(especially $B$, filter type and $N$) and examine the various trade-offs. This
is possible with our tool. \fig{fig:Redmill} shows the results of such an 
experiment where, with respect to the setting of~\fig{fig:Redmill_wordlengths}, 
we additionally allow $N$ to vary and also consider type II filters. 
We also added the flexible gain, genetic algorithm results produced in~\cite{rbd00}.
Compared to~\cite{rbd00}, we could improve all of the results, except two of them
where we obtained the same adder cost ($-9$ dB and $-25$ dB). It is clearly visible
that allowing variable gain designs can have a major influence on the quality of the results.



\begin{table*}[]
		\caption{Comparison between our method and the state-of-the-art results for the specifications in Table~\ref{tab:spec_reference_filters}. }\label{tab:results}
	\begin{tabular}{@{}cccccc>{\boldmath\textbf}cclclp{6.8cm}}
		\toprule
    Name & Source       & $N$   & Type & $A_\text{M}$ & $A_\text{S}$ & $A$   &  AD     & $G$            & $B$  & \makecell{Error} & Coefficients \\ 
    \midrule
    S1a  & \cite{s89}   & $24$  & I    & $11$           & $24$       & $35$  &   $2$   & $2.41$        & $8$  & $0.00159$   & $1$ $3$ $-1$ $-8$ $-7$ $10$ $20$ $-1$ $-40$ $-34$ $56$ $184$ $246$\\
    S1a  & ours         & $24$  & I    & $7$            & $20$       & $27$  &   $2$   & $1.251$        & $9$  & $0$         & $1$ $4$ $0$ $-8$ $-7$ $10$ $22$ $0$ $-41$ $-36$ $57$ $192$ $256$ \\ 
    S1a  & ours         & $24$  & I    & $6$            & $20$       & $26$  &   $2$   & $1.245678$     & $9$  & $0.00159$   & $1$ $4$ $0$ $-8$ $-8$ $10$ $22$ $0$ $-40$ $-37$ $57$ $192$ $256$ \\
    S1a  & ours         & $23$  & II   & $7$            & $19$       & $26$  &   $2$   & $2.654716$     & $8$  & $0$         & $3$ $3$ $-5$ $-11$ $0$ $20$ $16$ $-23$ $-52$ $0$ $134$ $253$ \\
    S1a  & ours         & $23$  & II   & $5$            & $19$       & $24$  &   $2$   & $2.172388$     & $8$  & $0.00159$   & $2$ $2$ $-3$ $-9$ $0$ $16$ $13$ $-18$ $-42$ $0$ $110$ $208$ \\ 
    \midrule
    S1b  & \cite{rbd00} & $24$  & I    & $6$            & $20$       & $26$  &   $3$   & $2.4570$       & $9$ & $0$         & $2$ $8$ $0$ $-16$ $-14$ $20$ $43$ $0$ $-80$ $-71$ $112$ $377$ $502$\\
    S1b  & ours         & $24$  & I    & $6$            & $20$       & $26$  &   $2$   & $1.40946$      & $9$ & $0$         & $2$ $4$ $0$ $-10$ $-8$ $12$ $24$ $0$ $-47$ $-40$ $65$ $216$ $288$\\ 
    S1b  & ours         & $23$  & II   & $5$            & $19$       & $24$  &   $2$   & $2.46492$      & $9$ & $0$         & $6$ $6$ $-8$ $-21$ $0$ $36$ $32$ $-42$ $-96$ $0$ $248$ $472$ \\     
    S1b  & ours         & $23$  & II   & $7$            & $19$       & $26$  &   $2$   & $2.65462$      & $8$  & $0$         & $3$ $3$ $-5$ $-11$ $0$ $20$ $16$ $-23$ $-52$ $0$ $134$ $253$ \\ 
    \midrule
    S1c  & \cite{yl07}  & $24$  & I    & $4$            & $24$       & $28$  &   $2$   & $1.8950$       & $8$  & $0$         & $2$ $3$ $-2$ $-8$ $-4$ $10$ $16$ $-3$ $-32$ $-24$ $48$ $144$ $191$\\
    S1c  & ours         & $24$  & I    & $5$            & $20$       & $25$  &   $2$   & $1.25615$      & $8$  & $0$         & $1$ $2$ $0$ $-4$ $-3$ $6$ $11$ $0$ $-21$ $-18$ $29$ $96$ $128$\\ 
    S1c  & ours         & $23$  & II   & $5$            & $19$       & $24$  &   $2$   & $1.86904$      & $7$  & $0$         & $1$ $1$ $-2$ $-4$ $0$ $7$ $6$ $-8$ $-18$ $0$ $47$ $89$ \\
    \cmidrule(rl){1-12}
    S1c  & \cite{sy11a} & $23$  & II   & $4$            & $19$       & $23$  &   $2$   & $1.34766$      & $8$                & $0.00118$   & $2$ $2$ $-2$ $-5$ $0$ $10$ $8$ $-12$ $-26$ $0$ $68$ $128$\\
    S1c  & ours         & $23$  & II   & $4$            & $19$       & $23$  &   $2$   & $1.34717$      & $8$                & $0.00118$   & $2$ $2$ $-2$ $-5$ $0$ $10$ $8$ $-12$ $-26$ $0$ $68$ $128$\\
    \midrule
    S2a  & \cite{s89}   & $59$  & II   & $57$           & $59$       & $116$ &   $2$   & $7.1324$       & $13$ & $0$         & $31$ $28$ $29$ $22$ $8$ $-17$ $-59$ $-116$ $-188$ $-268$ $-352$ $-432$ $-500$ $-532$ $-529$ $-464$ $-336$ $-129$ $158$ $526$ $964$ $1472$ $2008$ $2576$ $3136$ $3648$ $4110$ $4478$ $4737$ $4868$ \\
    S2a  & ours         & $59$  & II   & $22$           & $59$       & $81$  &   $2$   & $9.25424$      & $10$ & $0$         & $4$ $4$ $4$ $4$ $1$ $-2$ $-9$ $-18$ $-30$ $-42$ $-56$ $-69$ $-80$ $-86$ $-85$ $-76$ $-56$ $-22$ $24$ $84$ $155$ $236$ $325$ $416$ $508$ $593$ $668$ $728$ $770$ $792$\\ 
    \midrule
    S2b  & \cite{yl07}  & $59$  & II   & $19$           & $59$       & $78$  &   $3$   & $10.6888$       & $10$ & $0$         & $5$ $5$ $6$ $5$ $3$ $-2$ $-10$ $-20$ $-32$ $-48$ $-64$ $-80$ $-91$ $-99$ $-99$ $-88$ $-64$ $-26$ $28$ $96$ $178$ $273$ $376$ $482$ $587$ $686$ $772$ $842$ $892$ $916$\\
    S2b  & \cite{yl07}  & $59$  & II   & $21$           & $59$       & $80$  &   $2$   & $10.48712$      & $10$ & $0$         & $5$ $5$ $5$ $4$ $2$ $-4$ $-10$ $-20$ $-34$ $-48$ $-64$ $-78$ $-91$ $-98$ $-96$ $-86$ $-62$ $-24$ $28$ $96$ $176$ $269$ $369$ $473$ $575$ $672$ $756$ $824$ $872$ $897$\\
    S2b  & ours         & $59$  & II   & $19$           & $57$       & $76$  &   $2$   & $10.506472$     & $10$ & $0$         & $4$ $4$ $5$ $4$ $0$ $-4$ $-11$ $-22$ $-34$ $-49$ $-64$ $-79$ $-90$ $-98$ $-96$ $-84$ $-60$ $-24$ $30$ $97$ $178$ $270$ $370$ $474$ $576$ $672$ $756$ $824$ $872$ $896$\\ 
    \cmidrule(rl){1-12}
    S2b  & \cite{sy11a} & $59$  & II   & $17$           & $59$       & $76$  &   $3$   & $10.47032$      & $10$ & $0.01395$   & $5$ $5$ $6$ $5$ $2$ $-2$ $-10$ $-20$ $-32$ $-48$ $-64$ $-78$ $-92$ $-98$ $-87$ $-65$ $-26$ $26$ $93$ $174$ $267$ $368$ $472$ $575$ $672$ $757$ $826$ $874$ $898$\\
    S2b  & ours         & $59$  & II   & $15$           & $51$       & $66$  &   $2$   & $7.5904$       & $10$ & $0.00789$   & $0$ $0$ $0$ $-2$ $-5$ $-10$ $-16$ $-23$ $-32$ $-40$ $-50$ $-58$ $-64$ $-64$ $-61$ $-50$ $-29$ $0$ $38$ $86$ $143$ $206$ $274$ $344$ $412$ $476$ $532$ $576$ $608$ $624$\\ 
    \midrule
    L2   & \cite{yl07}  & $62$  & I    & $17$           & $62$       & $79$  &   $3$   & $2.6668$       & $11$ & $0$         & $4$ $9$ $13$ $12$ $4$ $-10$ $-26$ $-36$ $-32$ $-12$ $18$ $44$ $52$ $32$ $-10$ $-56$ $-80$ $-64$ $-4$ $74$ $130$ $128$ $48$ $-86$ $-215$ $-263$ $-168$ $88$ $460$ $854$ $1153$ $1265$\\
    L2   & ours         & $62$  & I    & $16$           & $62$       & $78$  &   $3$   & $2.6668$       & $11$ & $0$         & $4$ $9$ $13$ $12$ $4$ $-10$ $-26$ $-36$ $-32$ $-12$ $18$ $44$ $52$ $32$ $-10$ $-56$ $-80$ $-64$ $-4$ $74$ $130$ $128$ $48$ $-86$ $-215$ $-263$ $-168$ $88$ $460$ $854$ $1153$ $1265$ \\ 
    \midrule
    L3   & \cite{yl07}  & $35$  & II   & $3$            & $35$       & $38$  &   $2$   & $3.192$        & $8$  & $0$         & $8$ $1$ $-6$ $-12$ $-10$ $-1$ $6$ $20$ $20$ $6$ $-12$ $-32$ $-40$ $-16$ $32$ $96$ $160$ $196$\\
    L3   & ours         & $35$  & II   & $5$            & $33$       & $38$  &   $2$   & $2.58268$      & $7$  & $0$         & $4$ $0$ $-2$ $-4$ $-4$ $-1$ $4$ $7$ $8$ $3$ $-5$ $-14$ $-16$ $-7$ $12$ $39$ $64$ $79$\\
    L3   & ours         & $35$  & II   & $5$            & $31$       & $36$  &   $1$   & $2.6257$      & $8$  & $0$         & $7$ $0$ $-5$ $-8$ $-10$ $0$ $6$ $15$ $15$ $8$ $-12$ $-28$ $-32$ $-14$ $24$ $80$ $130$ $160$\\
    L3   & ours         & $35$  & II   & $4$            & $31$       & $35$  &   $2$   & $2.10468$      & $8$  & $0$         & $5$ $0$ $-4$ $-8$ $-8$ $0$ $4$ $11$ $13$ $5$ $-10$ $-22$ $-26$ $-11$ $20$ $64$ $104$ $129$\\
    \cmidrule(rl){1-12}
    L3   & \cite{sy11a} & $35$  & II   & $4$            & $31$       & $35$  &   $1$   & $2.627$       & $7$  & $0.00213$   & $3$ $0$ $-2$ $-5$ $-5$ $0$ $3$ $7$ $8$ $3$ $-6$ $-14$ $-16$ $-7$ $12$ $40$ $65$ $80$\\
    L3   & ours         & $35$  & II   & $4$            & $31$       & $35$  &   $1$   & $2.61998$      & $7$  & $0.00213$   & $3$ $0$ $-2$ $-4$ $-6$ $0$ $3$ $7$ $7$ $4$ $-6$ $-14$ $-16$ $-7$ $12$ $40$ $65$ $80$\\
    L3   & ours         & $35$  & II   & $5$            & $29$       & $34$  &   $2$   & $2.6211$      & $8$  & $0.00213$   & $7$ $0$ $0$ $-10$ $-8$ $0$ $8$ $16$ $16$ $7$ $-10$ $-28$ $-32$ $-14$ $24$ $78$ $130$ $160$\\
    L3   & ours         & $35$  & II   & $3$            & $31$       & $34$  &   $1$   & $2.60028$      & $8$  & $0.00213$   & $6$ $0$ $-3$ $-10$ $-10$ $0$ $7$ $12$ $16$ $7$ $-12$ $-28$ $-32$ $-14$ $24$ $80$ $128$ $160$\\
    L3   & ours         & $35$  & II   & $3$            & $31$       & $34$  &   $2$   & $2.60564$      & $7$  & $0.00213$   & $4$ $0$ $-2$ $-4$ $-4$ $0$ $3$ $8$ $8$ $3$ $-6$ $-13$ $-16$ $-8$ $13$ $40$ $64$ $80$\\    
    \bottomrule
	\end{tabular}
\end{table*}

\subsubsection{The Set of State-of-the-art Specifications}

Table~\ref{tab:results} presents the comparison between our designs and the best results 
from literature~\cite{lp83,s89,rbd00,yl07,ysl09,sy11a,szlm12} for the filter specifications 
from Table~\ref{tab:spec_reference_filters}. The following information for each implementation 
is given: filter order ($N$), filter type, number of multiplier adders ($A_M$), number of 
structural adders ($A_S$), total number of adders ($A$), adder depth of the multiplier block
(AD), gain ($G$), effective coefficient word length ($B$), frequency response error and, 
finally, coefficients of the filter. The frequency response error represents by 
how much the resulting filter does not respect the specification (zero in case of no issue)
and was necessary to introduce since coefficients from previous work turned out to sometimes 
slightly violate the specification. For those cases we {also} adjusted the specification 
to find solutions with a similar error {in order} to perform a fair {and complete} 
comparison. In the following, we discuss each instance in detail.

\paragraph*{S1a} we obtain a 23\% improvement (27 vs. 35 adders) in comparison to the implementation 
in~\cite{s89}. By also considering type II designs, one extra adder can be saved (26 total adders). If 
furthermore a slight error is allowed, a design with 24 adders is achievable.

\paragraph*{S1b} for this specification we show that, for type I designs, the AD can be 
reduced from 3 to 2 stages, while keeping the same effective word length. If we use type II filters,
keeping the same word length as~\cite{rbd00} allows the reduction of two adders (one structural and one MB). 
We can also decrease the word length
by one, and still get a result with the same cost as~\cite{rbd00}.

\paragraph*{S1c} we improve the result from~\cite{yl07} by 3 adders {by prioritizing} sparse implementations.
{For type II specifications we can gain one extra structural adder and decrease the word length by one.} {A} type II implementation 
of these specifications from~\cite{sy11a} requires only 23 adders but {has higher error than that allowed by} the specification. 
{If we allow 
some error, we obtain the same 23 adder solution as~\cite{sy11a}.} 

\paragraph*{S2a} we offer a significant improvement in the implementation of this specification 
w.r.t. the best known {result}~\cite{s89}. Our solution requires 30\% 
less adders (81 adders vs. 116 adders) and we reduce the {effective} word length from 13 
to 10 bits. 

\paragraph*{S2b} in~\cite{yl07}, {implementations with adder depths 3 and 2 are proposed,} at the cost of 78 and 80 adders, 
respectively. These results are improved in~\cite{sy11a}, with the authors claiming that a 
3-stage implementation at the cost of 76 adders has a high probability to be optimal. Our check 
show{s} that the latter design does not pass an a posteriori 
validation by a margin of at least $0.01395$. If some error margin is indeed acceptable, 
we demonstrate that a 2-stage design with a cost of only 66 adders is possible. Again, our 
result has higher sparsity than {previous} designs. If frequency 
specifications must be rigorously met, we show that 76 adders is the optimal cost for a 
2-stage design. 

\paragraph*{L1} the tool timed out for this specification {before giving a feasible
result}, hence it is not presented in Table~\ref{tab:results}. The best known 
result from literature is a 120-tap filter~\cite{yl07} and the size of an instance of the 
corresponding ILP formulation goes beyond the current capabilities of the solvers we tried, 
showing its limitations.

\paragraph*{L2} the result of~\cite{yl07} can be improved by one adder with the same 
coefficient set. Similar to \emph{L1}, we had to timeout before the solver could ascertain 
if the feasible filter obtained is indeed optimal.

\paragraph*{L3} in~\cite{yl07}, an implementation with 38 adders is provided. As in the case 
of the S2b specification,~\cite{sy11a} reduces both the adder count and adder depth, but at 
the cost of a substantial violation of the frequency specifications. We {improve on} the result 
from~\cite{yl07}, proving that 
{only a 7-bit wordlength is necessary 
to get the same 38 adder cost.} This cost can be 
reduced to 35 adders by {using 8}-bit coefficients. {If we allow the same error as 
in~\cite{sy11a}, we show that 34 adder solutions are possible.}

Overall, the proposed tool achieves significant improvements to the majority of the considered 
filter design problems. Morover, the user can explore a large design space by setting different 
implementation parameters, \emph{e.g.} adder depth, coefficient word length, filter type, etc. 
The required runtime however, will depend greatly on the problem. For producing
the results in Table~\ref{tab:results}, it varied from several seconds for the smallest filters
(S1) up to several days for the largest ones (S2 and L2).

\section{Conclusion}\label{sect:conclusion}
In this paper we have introduced two new algorithms for the
design of optimal multiplierless FIR filters. Relying on ILP
formulations stemming from the MCM literature, our algorithms
minimize either (a) the number of structural adders given a
fixed budget of multiplier block adders or (b) the total number
of adders (multiplier block + structural adders) given a fixed
adder depth. We further show how (a) can be applied iteratively 
to optimally minimize the total number of adders 
(without any adder count or adder depth constraints). 
Extensive numerical tests with example design problems 
from the state-of-the-art show that our 
approaches can offer in many cases better results. We also make 
available an open-source C++ implementation of the proposed
methods.

\normalem

\bibliographystyle{IEEEtran}
\bibliography{IEEEabrv,fir_ilp}

\begin{thebibliography}{10}
\providecommand{\url}[1]{#1}
\csname url@samestyle\endcsname
\providecommand{\newblock}{\relax}
\providecommand{\bibinfo}[2]{#2}
\providecommand{\BIBentrySTDinterwordspacing}{\spaceskip=0pt\relax}
\providecommand{\BIBentryALTinterwordstretchfactor}{4}
\providecommand{\BIBentryALTinterwordspacing}{\spaceskip=\fontdimen2\font plus
\BIBentryALTinterwordstretchfactor\fontdimen3\font minus
  \fontdimen4\font\relax}
\providecommand{\BIBforeignlanguage}[2]{{%
\expandafter\ifx\csname l@#1\endcsname\relax
\typeout{** WARNING: IEEEtran.bst: No hyphenation pattern has been}%
\typeout{** loaded for the language `#1'. Using the pattern for}%
\typeout{** the default language instead.}%
\else
\language=\csname l@#1\endcsname
\fi
#2}}
\providecommand{\BIBdecl}{\relax}
\BIBdecl

\bibitem{lp83}
Y.~Lim and S.~Parker, ``{FIR filter design over a discrete powers-of-two
  coefficient space},'' \emph{IEEE Transactions on Acoustics, Speech, and
  Signal Processing}, vol.~31, no.~3, pp. 583--591, 1983.

\bibitem{s89}
H.~Samueli, ``{An improved search algorithm for the design of multiplierless
  FIR filters with powers-of-two coefficients},'' \emph{IEEE Transactions on
  Circuits and Systems}, vol.~36, no.~7, pp. 1044--1047, Jul. 1989.

\bibitem{h96}
R.~I. Hartley, ``Subexpression sharing in filters using canonic signed digit
  multipliers,'' \emph{IEEE Transactions on Circuits and Systems II: Analog and
  Digital Signal Processing}, vol.~43, no.~10, pp. 677--688, 1996.

\bibitem{rbd00}
D.~Redmill, D.~Bull, and E.~Dagless, ``{Genetic synthesis of reduced complexity
  filters and filter banks using primitive operator directed graphs},''
  \emph{Circuits, Devices and Systems, IEE Proceedings -}, vol. 147, no.~5, pp.
  303--310, 2000.

\bibitem{gj01}
O.~Gustafsson and H.~Johansson, ``{An MILP approach for the design of
  linear-phase FIR filters with minimum number of signed-power-of-two terms},''
  \emph{es.isy.liu.se}, 2001.

\bibitem{ys01}
J.~Yli-Kaakinen and T.~Saramaki, ``{A systematic algorithm for the design of
  multiplierless FIR filters},'' \emph{IEEE International Symposium on Circuits
  and Systems}, vol.~2, pp. 185--188, 2001.

\bibitem{gw02}
O.~Gustafsson and L.~Wanhammar, ``{Design of linear-phase FIR filters combining
  subexpression sharing with MILP},'' in \emph{Midwest Symposium on Circuits
  and Systems}.\hskip 1em plus 0.5em minus 0.4em\relax IEEE, Aug. 2002, pp.
  III--9--III--12.

\bibitem{v05}
A.~P. Vinod and E.-K. Lai, ``On the implementation of efficient channel filters
  for wideband receivers by optimizing common subexpression elimination
  methods,'' \emph{IEEE Transactions on Computer-Aided Design of Integrated
  Circuits and Systems}, vol.~24, no.~2, pp. 295--304, 2005.

\bibitem{yl07}
Y.~J. Yu and Y.~C. Lim, ``{Design of Linear Phase FIR Filters in Subexpression
  Space Using Mixed Integer Linear Programming},'' \emph{IEEE Transactions on
  Circuits and Systems I: Regular Papers}, vol.~54, no.~10, pp. 2330--2338,
  2007.

\bibitem{a08}
M.~Aktan, A.~Yurdakul, and G.~Dundar, ``{An algorithm for the design of
  low-power hardware-efficient FIR filters},'' \emph{IEEE Transactions on
  Circuits and Systems I: Regular Papers}, vol.~55, no.~6, pp. 1536--1545,
  2008.

\bibitem{yl09}
Y.~Yu and Y.~Lim, ``{Optimization of Linear Phase FIR Filters in Dynamically
  Expanding Subexpression Space},'' \emph{Circuits, Systems, and Signal
  Processing}, vol.~29, no.~1, pp. 1--16, 2009.

\bibitem{ysl09}
Y.~J. Yu, D.~Shi, and Y.~C. Lim, ``{Design of extrapolated impulse response FIR
  filters with residual compensation in subexpression space},'' \emph{IEEE
  Transactions on Circuits and Systems I: Regular Papers}, 2009.

\bibitem{sy11a}
D.~Shi and Y.~J. Yu, ``{Design of Linear Phase FIR Filters With High
  Probability of Achieving Minimum Number of Adders},'' \emph{IEEE Transactions
  on Circuits and Systems I: Regular Papers}, vol.~58, no.~1, pp. 126--136,
  2011.

\bibitem{sy11b}
------, ``{Design of Discrete-Valued Linear Phase FIR Filters in Cascade
  Form},'' \emph{IEEE Transactions on Circuits and Systems I: Regular Papers},
  vol.~58, no.~7, pp. 1627--1636, 2011.

\bibitem{szlm12}
A.~Shahein, Q.~Zhang, N.~Lotze, and Y.~Manoli, ``{A Novel Hybrid Monotonic
  Local Search Algorithm for FIR Filter Coefficients Optimization},''
  \emph{IEEE Transactions on Circuits and Systems I: Regular Papers}, vol.~59,
  no.~3, pp. 616--627, 2012.

\bibitem{by14}
W.~Bin~Ye and Y.~J. Yu, ``{A polynomial-time algorithm for the design of
  multiplierless linear-phase FIR filters with low hardware cost},'' \emph{IEEE
  International Symposium on Circuits and Systems (ISCAS)}, pp. 970--973, 2014.

\bibitem{ye14}
W.~B. Ye and Y.~J. Yu, ``{Bit-level multiplierless FIR filter optimization
  incorporating sparse filter technique},'' \emph{IEEE Transactions on Circuits
  and Systems I: Regular Papers}, vol.~61, no.~11, pp. 3206--3215, 2014.

\bibitem{mw15}
A.~Mehrnia and A.~N. Willson, ``{Optimal Factoring of FIR Filters},''
  \emph{IEEE Transactions on Signal Processing}, vol.~63, no.~3, pp. 647--661,
  Feb. 2015.

\bibitem{yly17}
W.~B. Ye, X.~Lou, and Y.~J. Yu, ``{Design of Low Power Multiplierless
  Linear-Phase FIR Filters},'' \emph{IEEE Access}, 2017.

\bibitem{lmyy17}
X.~Lou, P.~K. Meher, Y.~Yu, and W.~Ye, ``{Novel Structure for Area-Efficient
  Implementation of FIR Filters},'' \emph{IEEE Transactions on Circuits and
  Systems II: Express Briefs}, vol.~64, no.~10, pp. 1212--1216, 2017.

\bibitem{cs84}
P.~Cappello and K.~Steiglitz, ``{Some Complexity Issues in Digital Signal
  Processing},'' \emph{IEEE Transactions on Acoustics, Speech and Signal
  Processing}, vol.~32, no.~5, pp. 1037--1041, Oct. 1984.

\bibitem{dm95a}
A.~Dempster and M.~Macleod, ``{Use of Minimum-Adder Multiplier Blocks in FIR
  Digital Filters},'' \emph{IEEE Transactions on Circuits and Systems II:
  Analog and Digital Signal Processing}, vol.~42, no.~9, pp. 569--577, 1995.

\bibitem{vp07}
Y.~Voronenko and M.~P{\"u}schel, ``{Multiplierless Multiple Constant
  Multiplication},'' \emph{ACM Transactions on Algorithms}, vol.~3, no.~2, pp.
  1--38, 2007.

\bibitem{g07a}
O.~Gustafsson, ``{A Difference Based Adder Graph Heuristic for Multiple
  Constant Multiplication Problems},'' in \emph{IEEE International Symposium on
  Circuits and Systems (ISCAS)}, 2007, pp. 1097--1100.

\bibitem{g08}
------, ``{Towards Optimal Multiple Constant Multiplication: A Hypergraph
  Approach},'' in \emph{Asilomar Conference on Signals, Systems and Computers
  (ACSSC)}.\hskip 1em plus 0.5em minus 0.4em\relax IEEE, Oct. 2008, pp.
  1805--1809.

\bibitem{agf10}
L.~Aksoy, E.~G{\"u}nes, and P.~Flores, ``{Search Algorithms for the Multiple
  Constant Multiplications Problem: Exact and Approximate},''
  \emph{Microprocessors and Microsystems}, vol.~34, no.~5, pp. 151--162, 2010.

\bibitem{k18}
M.~Kumm, ``{Optimal Constant Multiplication using Integer Linear
  Programming},'' \emph{IEEE Transactions on Circuits and Systems II: Express
  Briefs}, 2018.

\bibitem{co75}
R.~E. Crochiere and A.~V. Oppenheim, ``{Analysis of Linear Digital Networks},''
  in \emph{Proceedings of the IEEE}, 1975, pp. 581--595.

\bibitem{k80}
D.~M. Kodek, ``{Design of optimal finite wordlength FIR digital filters using
  integer programming techniques},'' \emph{IEEE Transactions on Acoustics,
  Speech and Signal Processing}, vol.~28, no.~3, pp. 304--308, 1980.

\bibitem{de83}
J.~M. de~Sa, ``{A new design method of optimal finite wordlength linear phase
  FIR digital filters},'' \emph{IEEE Transactions on Acoustics, Speech, and
  Signal Processing}, vol.~31, no.~4, pp. 1032--1034, 1983.

\bibitem{l90}
Y.~Lim, ``{Design of discrete-coefficient-value linear phase FIR filters with
  optimum normalized peak ripple magnitude},'' \emph{IEEE Transactions on
  Circuits and Systems}, vol.~37, no.~12, pp. 1480--1486, 1990.

\bibitem{k99}
D.~M. Kodek, ``Design of optimal finite wordlength {FIR} digital filters,'' in
  \emph{Proc. of the European Conference on Circuit Theory and Design, ECCTD
  '99.}, vol.~1, Aug. 1999, pp. 401--404.

\bibitem{k05}
------, ``{Performance limit of finite wordlength FIR digital filters},''
  \emph{IEEE Transactions on Signal Processing}, vol.~53, no.~7, pp.
  2462--2469, 2005.

\bibitem{k12}
------, ``{LLL algorithm and the optimal finite wordlength FIR design},''
  \emph{IEEE Transactions on Signal Processing}, vol.~60, no.~3, pp.
  1493--1498, 2012.

\bibitem{Antoniou2005}
A.~Antoniou, \emph{{Digital Signal Processing: Signals, Systems, and
  Filters}}.\hskip 1em plus 0.5em minus 0.4em\relax McGraw-Hill Education,
  2005.

\bibitem{fc10}
M.~Faust and C.-H. Chang, ``{Minimal Logic Depth Adder Tree Optimization for
  Multiple Constant Multiplication},'' \emph{IEEE International Symposium on
  Circuits and Systems (ISCAS)}, pp. 457--460, 2010.

\bibitem{ddk00}
S.~Demirsoy, A.~Dempster, and I.~Kale, ``{Transition Analysis on FPGA for
  Multiplier-Block Based FIR Filter Structures},'' \emph{IEEE International
  Symposium of Circuits and Systems (ISCAS)}, vol.~2, pp. 862--865 vol.2, 2000.

\bibitem{ddk02b}
------, ``{Power Analysis of Multiplier Blocks},'' in \emph{IEEE International
  Symposium on Circuits and Systems (ISCAS)}, 2002, pp. I--297--I--300 vol.1.

\bibitem{ddk02a}
A.~G. Dempster, S.~S. Demirsoy, and I.~Kale, ``{Designing Multiplier Blocks
  with Low Logic Depth},'' in \emph{IEEE International Symposium on Circuits
  and Systems (ISCAS)}.\hskip 1em plus 0.5em minus 0.4em\relax IEEE, 2002, pp.
  V--773--V--776.

\bibitem{k15}
M.~Kumm, ``{Multiple Constant Multiplication Optimizations for Field
  Programmable Gate Arrays},'' Ph.D. dissertation, Springer Wiesbaden,
  Wiesbaden, Oct. 2015.

\bibitem{dm94}
A.~Dempster and M.~D. Macleod, ``{Constant Integer Multiplication Using Minimum
  Adders},'' \emph{IEE Proceedings of Circuits, Devices and Systems}, vol. 141,
  no.~5, pp. 407--413, 1994.

\bibitem{kfmzm13}
M.~Kumm, D.~Fangh{\"a}nel, K.~M{\"o}ller, P.~Zipf, and U.~Meyer-Baese, ``{FIR
  Filter Optimization for Video Processing on FPGAs},'' \emph{Springer EURASIP
  Journal on Advances in Signal Processing}, pp. 1--18, 2013.

\bibitem{kp01}
H.-J. Kang and I.-C. Park, ``{FIR Filter Synthesis Algorithms for Minimizing
  the Delay and the Number of Adders},'' \emph{IEEE Transactions on Circuits
  and Systems II: Analog and Digital Signal Processing}, vol.~48, no.~8, pp.
  770--777, 2001.

\bibitem{cr73}
D.~Chan and L.~Rabiner, ``{Analysis of Quantization Errors in the Direct Form
  for Finite Impulse Response Digital Filters},'' \emph{IEEE Transactions on
  Audio and Electroacoustics}, vol.~21, no.~4, pp. 354--366, Aug. 1973.

\bibitem{Volk17c}
A.~Volkova, C.~Lauter, and T.~Hilaire, ``Reliable verification of digital
  implemented filters against frequency specifications,'' in \emph{2017 IEEE
  24th Symposium on Computer Arithmetic (ARITH)}, July 2017, pp. 180--187.

\bibitem{f16}
S.-I. Filip, ``{A robust and scalable implementation of the Parks-McClellan
  algorithm for designing FIR filters},'' \emph{ACM Transactions on
  Mathematical Software (TOMS)}, vol.~43, no.~1, p.~7, 2016.

\bibitem{bfh18}
N.~Brisebarre, S.-I. Filip, and G.~Hanrot, ``{A Lattice Basis Reduction
  Approach for the Design of Finite Wordlength FIR Filters},'' \emph{IEEE
  Transactions on Signal Processing}, vol.~66, no.~10, pp. 2673--2684, May
  2018.

\bibitem{os14}
A.~V. Oppenheim and R.~W. Schafer, \emph{Discrete-time signal
  processing}.\hskip 1em plus 0.5em minus 0.4em\relax Pearson Education, 2014.

\bibitem{scip}
A.~Gleixner \emph{et~al.}, ``{The SCIP Optimization Suite 6.0},''
  \url{http://www.optimization-online.org/DB_HTML/2018/07/6692.html},
  Optimization Online, Technical Report, July 2018.

\bibitem{gurobi}
``{Gurobi} optimization system,'' \url{https://www.gurobi.com}.

\bibitem{cplex}
``{CPLEX optimizer},'' \url{https://www.ibm.com/analytics/cplex-optimizer}.

\bibitem{sskz18}
P.~Sittel, T.~Sch{\"o}nw{\"a}lder, M.~Kumm, and P.~Zipf, ``{ScaLP: A
  Light-Weighted (MI)LP Library},'' in \emph{Methoden und Beschreibungssprachen
  zur Modellierung und Verifikation von Schaltungen und Systemen (MBMV)}, 2018,
  pp. 1--10.

\bibitem{k13}
D.~M. Kodek, ``{Length limit of optimal finite wordlength FIR filters},''
  \emph{Digital Signal Processing}, vol.~23, no.~5, pp. 1798--1805, Sep. 2013.

\end{thebibliography}

\end{document}